\documentclass[aps,pre,preprint,showpacs,groupedaddress,amsmath,amssymb]{revtex4}

\usepackage{graphicx}

\begin{document}


\title{Sonochemical effects on single-bubble sonoluminescence}


\author{Li Yuan}
\email{lyuan@lsec.cc.ac.cn}
\affiliation{LSEC and Institute of Computational Mathematics,
Academy of Mathematics and Systems Science, Chinese Academy of
Sciences, Beijing 100080, People's Republic of China}


\date{\today}

\begin{abstract}
A refined hydrochemical model for single-bubble sonoluminescence
(SBSL) is presented. The processes of water vapor evaporation and
condensation, mass diffusion, and chemical reactions are taken
into account. Numerical simulations of Xe-, Ar- and He-filled
bubbles are carried out. The results show that the trapped water
vapor in conjunction with its endothermic chemical reactions
significantly reduces the temperature within the bubble so that
the degrees of ionization are generally very low. The chemical
radicals generated from water vapor are shown to play an
increasingly important role in the light emission from Xe to He
bubbles. Light spectra and pulses are then computed from an
optically thin model. It is found that the resulting spectrum
intensities are too small and the pulse widths are too short to
fit to recent experimental results within stable SBSL range.
Addition of a finite-size blackbody core to the optically thin
model improves the fitting. Suggestions on how to reconcile the
conflict are given.
\end{abstract}


\maketitle

\section{introduction}

The discovery that acoustic energy can be converted to light
through an oscillating air bubble trapped in water \cite{Gait92}
has triggered extensive studies on single-bubble sonoluminescence
(SBSL) (see reviews in Refs.~\cite{Barber97,Bren02}). Under
certain conditions, a stable and regular flash of blue-white light
with a width of about 40$-$350~ps is emitted during the violent
collapse of the bubble in each acoustic cycle \cite{Gom,Hil,Mor}.
Among many candidate mechanisms of SBSL light emission, the model
that identified SBSL light emission as thermal bremsstrahlung and
recombination radiation from an optically thin bubble seemed to
predict the widths, shapes, and spectra of the emitted light
fairly well under certain simplified hydrodynamic frameworks
\cite{Hilg_nature,HilgPHF99,Ham00,Ham01}. Similar models included
those which used finite opacity to define a variable blackbody
core \cite{Moss97,Moss99,Burn01,Ho01,Ho02}, whose suitability was
reflected in the good fitting by a smaller-than-bubble blackbody
\cite{Vaz01}. However, different approximations in modelling the
physical-chemical processes, and particularly, uncertainties under
the extreme conditions inside a sonoluminescing bubble, may give
diverse predictions that plague the validity of a light emission
model. Therefore, more realistic hydrodynamic-chemical modelling
and more critical tests of the light emission models under the
refined hydro-chemical framework are necessary.

Past studies delineated the important effects of diffusive
transport, surface tension, and compressibility of the surrounding
liquid on SBSL \cite{Vuong96,Yuan98,Cheng}. Earlier studies
\cite{Fuji80, Kama93} considered the influence of
evaporation-condensation phenomena on the bubble dynamics. Sochard
\emph{et al.}~\cite{Soch97} and Gong \emph{et al.}~\cite{Gon98}
coupled the bubble dynamics with the water vapor dissociations.
Recently, the study of sonochemsitry was extended from lower
temperature to higher temperature situations where SBSL emerges
\cite{Yasui97,Sus99}. Yasui \cite{Yasui97} presented a model of
SBSL that accounts for evaporation-condensation process at the
bubble interface and water vapor chemical reactions. It was later
stressed by Storey and Szeri \cite{Stor00} and others
\cite{Moss99,Tog00, Tog02,Xie03} that water vapor reduce the
temperatures inside the SL bubble significantly by reducing the
compression heating of the mixture and  through primarily
endothermic chemical reactions. In some of these models, spatial
uniformity of the bubble interior was assumed and inter-molecular
mass diffusion was not properly accounted for. As a consequence,
such models tended to underpredict the amount of trapped water
vapor during the rapid collapse.

In the full hydrochemical numerical study by Storey and Szeri
\cite{Stor00}, the consequences of water vapor inside strongly
forced argon bubbles were investigated in detail. The interaction
of nonlinearity of the volume oscillations, mass diffusion, and
nonequilibrium phase change at the bubble wall resulted in excess
water vapor trapped in the bubble during the violent collapse. The
amount of trapped water vapor was more than that predicted by the
simple model \cite{Yasui97}.   Akhatov \emph{et al.}~\cite{Akha01}
accounted for the occurrence of supercritical conditions of
condensation and studied laser-induced cavitation bubbles. The
effects of water vapor diffusion in different noble gas bubbles
were studied by Xu \emph{et al.} \cite{Xu03}, where shock waves
were found to occur only in Xe bubbles.  More recently, Toegel
\emph{et al.}~\cite{Tog02} studied the effects of the highly
compressed conditions of SL bubbles on chemical equilibrium
constants. They showed that high temperatures could be recovered
due to the suppressed water vapor dissociations. In spite of the
progress made, the direct consequence of sonochemistry on existing
popular light emission models is to be revealed in  a full
hydrodynamic model.

In this paper, we present a refined hydrodynamic model taking into
account the chemical reactions and ionizations of the noble gas
and water vapor mixture.   The model is an extension of our
previous ones \cite{Yuan98, Ho01,Ho02}. As done for a pure argon
bubble \cite{Ho02}, the  Navier-Stokes (NS) equations for the
multispecies gas mixture in the bubble interior are coupled with a
proper form of the Rayleigh-Plesset (RP) equation for the bubble
wall, including the effects of liquid compressibility and heat
transfer. The newly added  feature is that the nonequilibrium
processes of evaporation and condensation, species diffusion,
noble gas exchange between the bubble and the surrounding liquid,
and dissociations of water vapor and ionizations of atomic species
inside the bubble are all taken into account. The numerical scheme
for solving the hydrodynamic equations is modified to be a
semi-implicit one which allows for better numerical stability than
original explicit scheme in Ref.~\cite{Yuan98}.

Detailed formulae in Eulerian framework are given. Numerical
simulations are carried out for bubbles of He, Ar, or Xe gases.
The effects of sonochemistry on current light-emitting models  of
SBSL \cite{HilgPHF99,Moss99,Ho02} are studied in detail through
comparison with a calibrated experiment \cite{Vaz01}. The main
conclusions are: (i) the chemical reactions reduce the temperature
within the bubble to such an extent that the degrees of ionization
are generally very low; (ii) shock waves do not appear in He or Ar
bubbles in the stable SBSL regime, but can occur in Xe bubbles
only at higher driving pressures; (iii) chemical radicals
generated from the water vapor contribute dominantly to the light
emission of He bubbles; (iv) based on the computed photon
absorption coefficients, the light spectra and pulse widths
computed from the popular optically thin model can hardly be
fitted to the experimental ones. Addition of a dynamic blackbody
core to the optically thin model improves the fitting. Some
suggestions on how to improve the optically thin model are
outlined.

\section{Hydrodynamic model}
In this section, we extend our previous hydrodynamic model
\cite{Yuan98,Ho02} to include processes of evaporation and
condensation on the wall and chemical reactions inside the bubble.
The bubble is assumed to be spherically symmetric and is composed
of the mixture of noble gas, water vapor, and reaction products.
In addition to the NS and RP equations, the equations for the mass
concentration of the dissolved noble gas and for the temperature
of the surrounding water are also solved.

\subsection{Gas dynamics in the bubble}

\subsubsection{The NS equations}

The bubble is assumed to contain $N$-species gas mixture.
Nonequilibrium chemical reactions of the water vapor and
ionizations of the  monatomic species (Ar, H, and O) are
considered. The maximum ionization level is taken as 3 for a noble
gas, and 1 for H or O atom. For an Ar bubble, $N=13$: Ar, Ar$^+$,
Ar$^{2+}$, Ar$^{3+}$, H$_2$O, OH, H, H$^+$, O, O$^+$, H$_2$,
O$_2$, e$^-$. The dynamics inside the bubble is described by the
compressible NS equations, which represent the conservation of
mass, momentum and energy. They can be written into a
``conservative" form  in the spherical coordinates
\begin{eqnarray}
\label{NS_eq}
 \frac{\partial\mathbf{Q}}{\partial t}
+\frac{\partial \mathbf{F}}{\partial r}
=\mathbf{H}+\frac{1}{r^2}\frac{ \partial r^2
\mathbf{F}_\nu}{\partial r} +\mathbf{M}_\nu +\mathbf{S},
\label{ns}
\end{eqnarray}
with
\begin{eqnarray}
\nonumber
{\bf Q}=\left[\begin{array}{c}\rho_1 \\ \vdots \\ \rho_{_N} \\
 \rho u \\ \rho E  \end{array}\right], \quad
{\bf F}=\left[\begin{array}{c}
\rho_1 u \\ \vdots \\ \rho_{_N} u \\
 \rho u^2+P \\u(\rho E+ P)  \end{array}\right], \quad
{\bf H}=-\frac{2 u}{r}\left[\begin{array}{c}
\rho_1 \\ \vdots \\ \rho_{_N} \\
 \rho u \\ \rho E + P \end{array}\right],
 \end{eqnarray}
\begin{eqnarray}
{\bf F}_\nu=\left[\begin{array}{c} -J_1 \\ \vdots \\ -J_{_N} \\
\tau_{rr} \\ u\tau_{rr}+ q \end{array}\right], \quad
{\bf M}_\nu=\left[\begin{array}{c}  0 \\ \vdots \\   0 \\
-(\tau_{\theta \theta}+\tau_{\phi\phi})/r  \\ 0 \end{array}\right], \quad
{\bf S}=\left[\begin{array}{c} \dot{\omega}_1 \\ \vdots \\
\dot{\omega}_{_N} \\0 \\0  \end{array}\right],
\end{eqnarray}
and
\begin{eqnarray}
\nonumber
\rho=\sum^N_{i=1}\rho_i, \quad
E=e +\frac{u^2}{2}, \quad  e=\sum^N_{i=1} e_i, \quad
\tau_{rr}=-2\tau_{\theta\theta}=-2\tau_{\phi\phi}=
\frac{4\mu}{3}\left(\frac{\partial u}{\partial r}-\frac{u}{r}\right),\quad
\end{eqnarray}
\begin{eqnarray}
\label{Diffu_eq}
q=\lambda \frac{\partial T}{\partial r} -\sum^N_{i=1}J_i h_i,\quad
J_i=-\rho D_i^M \frac{M_i}{\bar{M}}
\left[\frac{\partial X_i}{\partial r}+ (X_i-Y_i)
\frac{\partial \ln P}{\partial r}   + K^T_i \frac{\partial \ln T}{
\partial r} \right], \quad
\end{eqnarray}
\begin{eqnarray}
\nonumber \quad  Y_i=\frac{\rho_i}{\rho},\quad
X_i={Y_i}\frac{\bar{M}}{M_i}, \quad
\bar{M}=\frac{1}{\sum^N_{i=1}\frac{Y_i}{M_i}},\quad
D_i^M=\frac{1-Y_i}{\sum_{j\ne i}^N X_j/D_{ij}}, \quad
h_i=e_i+\frac{P_i}{\rho_i},
\end{eqnarray}
where $\rho_i$ is density of species $i$, $\rho$ is the total
density of gas mixture, $P$ is the pressure, $T$ is the
temperature, $u$ is the mass averaged velocity, $e$ is the total
internal energy of the mixture, $e_i$ and $h_i$ are the internal
energy and enthalpy of species $i$, respectively, $P_i$ is the
partial pressure,  $\mu$  and $\lambda$ are the viscosity and the
thermal conductivity of the mixture, respectively, $Y_i$  and $
X_i$ are the mass fraction and  mole fraction of species  $i$,
respectively, $M_i$ is the molar mass of species $i$, $\bar{M}$ is
the mean molar mass of the mixture, $D_{ij}$ is the binary
diffusion coefficient between species $i$ and $j$,  $D_i^M$ is the
mean diffusion coefficient of species $i$ into the mixture,
$K_i^T$ is the thermal diffusion ratio,  $\tau_{rr}$ is the normal
stress, $J_i$ is the mass diffusion flux that must satisfy $
\sum^N_{i=1} J_i=0 $, and $ \dot{\omega}_i$ is the net mass
production rate due to chemical reactions and ionizations that
satisfies $ \sum^N_{i=1} \dot{\omega}_i=0 $. Use of the mean
diffusion coefficient is a practical approximation for
computational efficiency \cite{Stor00, Poist}. However, to ensure
global mass conservation, a correction diffusion flux $J^c_i=
\rho_i\sum^N_{i=1}D_i^M \frac{M_i}{\bar{M}} \left[\frac{\partial
X_i}{\partial r}+ (X_i-Y_i) \frac{\partial \ln P}{\partial r}   +
K^T_i \frac{\partial \ln T}{
\partial r} \right]$  is added to $J_i$  in Eq.(\ref{Diffu_eq}), as
recommended by Ref.~\cite{Poist}. It can be easily shown that the
modified  $ J_i$ satisfies $ \sum^N_{i=1} J_i=0$.

\subsubsection{Transport properties, equation of state and thermodynamic properties}

The individual transport properties (viscosity $\mu_i$, thermal
conductivity $\lambda_i$, thermal diffusion ratio $k^T_i$, and
binary diffusion coefficient $D_{ij}$) are generally calculated
based on Chapman-Enskog theory \cite{Hirsc54,Reid,Stor99}.
However, the transport properties of the gas mixture,  $\mu,
\lambda $ are determined by some empirical combination rules such
as Wilke's semiempirical formula \cite{Reid}. There is also
difficulty in describing individual $\mu_i$ and $D_{ij}$ for some
reaction products due to lack of data. We shall let unavailable
$\mu_i$ and $D_{ij}$ equal to other known ones, e.g., $\mu_{\rm
OH}= \mu_{\rm H_2O}, \mu_{\rm H}= \mu_{\rm H_2}$. Collision cross
sections of ions, electron, and some radical species that are not
available in Ref.~\cite{Hirsc54,Reid} can be determined using NASA
temperature-dependent polynomial fitting \cite{Gupta}. Once
$\mu_i$ is known, $\lambda_i$ is obtained by modified Eucken model
\cite{Hirsc54,Reid}. The trouble lies in the determination of
$k^T_i$. In this regard, we only take into account the thermal
diffusion between the noble gas and the water vapor, since thermal
diffusion is only important in slow stage other than during
collapse \cite{Stor99,Stor00}. The high pressure corrections
\cite{Reid} are applied to $\mu_i$ and $\lambda_i$ similar to what
Xu \cite{Xu03} did. The transport properties of the mixture $\mu$
and $\lambda$ are obtained by using Wilke's semiempirical formula
\begin{eqnarray}
\mu= \sum_{i=1}^N \left(\frac{ X_i \mu_i}{\sum_{j=1}^N X_j
\phi_{ij}}\right),
\\
\lambda= \sum_{i=1}^N \left(\frac{ X_i \lambda_i}{ \sum_{j=1}^N
X_j \phi_{ij}} \right),
\end{eqnarray}
where
\begin{eqnarray}
\phi_{ij} =  \left[1 + \left(\frac{\mu_i}{\mu_j}\right)^{1/2}
\left(\frac{M_j}{M_i}\right)^{1/4} \right]^2 \left[ \sqrt{8}
\left( 1+ \frac{M_i}{M_j}\right) ^{1/2} \right] ^{-1}.
\end{eqnarray}

The hydrodynamics of  the bubble is affected by the equation of
state. Here the gas mixture is modeled by a hardcore van der Waals
equation of state that has the excluded volume but ignores the van
der Waals force as the previous authors did \cite{Wu93}
\begin{eqnarray}
\label{veos} P=\sum^N_{i=1}P_i = \frac{ \rho R T}{1- b \rho}=P(T,
\rho_1,\cdots,\rho_{_N}) , \end{eqnarray} where
$R=\sum^N_{i=1}Y_iR_i,  R_i=R_u/M_i $,  $R_u$ is the universal gas
constant, $ b= \sum^N_{i=1}Y_ib_i$ is a simple combination of
$b_i$, with $b_i$ being the van der Waals excluded volume in
m$^3$/kg. The values of $\tilde{b}_i= b_i M_i $ (in cm$^3$/mol)
are computed by $ \tilde{b}=RT_c/8P_c$ where $T_c$ and $P_c$ are
critical temperature and pressure of a species \cite{Reid}. When
these critical parameters are not available, $b_i$ is taken as 4
times the spherical volume of the atomic or ionic radius.

A well-posedness of equation closure requires
$P=f(e,\rho_1,\cdots,\rho_{_N})$, therefore, $T$ must be solved
for from the given energy relation
\begin{eqnarray}
 E= e  +\frac{u^2}{2}, \qquad e= \sum^N_{i=1} Y_i \left(C_{V i}
R_i  T + e^0_i\right) + \sum_{i={\rm molecues}}Y_i e_i^V +
\sum_{i={\rm ions}} Y_i e_i^I,
\end{eqnarray}
where the internal energy is divided into the translational and
rotational, the vibrational, and the ionization parts. The
coefficient of specific heat at constant volume $C_{V i}$, is
approximately assumed as follows \cite{Yasui97}: $C_{V i}=\frac
32$ for monatomic gases such as Ar, Ar$^+$, H, O; $C_{V i}= \frac
52 $ for diatomic gases such as OH, H$_2$, O$_2$, and $C_{V i} =
\frac 6 2 $ for other gases. $e_i^0$ is the reference energy,
$e^V_i$ is the vibrational energy of molecules species, and
$e^I_i$ is the ionization energy of ion species. The reference
energy $e_i^0$ is taken to be standard heat of formation at 298 K
\cite{GRI3}.

\subsubsection{Chemical kinetics}

The chemical kinetics consists of the reaction mechanism and
determines the net production rate of each species. For the
chemical reactions of water vapor, we use the mechanism that were
described in detail in Ref.~\cite{Evans81}. Only a subset
consisting of eight elementary reactions is used, corresponding to
the first eight ones used by Yasui \cite{Yasui97}. The first 19
reactions of Ref.~\cite{Yasui97} with additional species (HO$_2$,
H$_2$O$_2$) were also tried but the resulting temperature was
found to be a little lower than that from the 8 reaction scheme.
The processes of nonequilibrium collisional ionization and
recombination \cite{Xu99} are considered only for three monatomic
species, noble gas, H,  and O. The reason  to choose H and O atoms
is that they are quite ample in the water vapor dissociations,
have lower ionization potentials, and can be treated using
previous rate formulae. Ionizations of molecular species such as
OH and H$_2$O are believed to be more complicated, thus are not
accounted for. The net mass production rate $\dot{\omega}_k$ due
to chemical reactions of the water vapor is determined by the law
of mass action
\begin{eqnarray}
\dot{\omega}_k &= &M_k\sum^{N_r}_{i=1} \left( \nu^{\prime\prime}_{ki}-
\nu^\prime_{ki} \right) q_i,  \\
\label{progress_eq}  q_i &= &  k_{f_i} \prod^N_{k=1}
\left(\frac{\rho_k}{M_k}\right)^{\nu^\prime_{ki}}
-k_{b_i}\prod^N_{k=1}\left(\frac{\rho_k}{M_k}\right)^{\nu^{\prime\prime}_{ki}},
\end{eqnarray}
where $q_i$ is the net rate of progress of reaction $i$, and $N_r$
the total number of reactions. The forward and backward reaction
rate constants for the $i$th reaction $ k_{f_i} $ and $ k_{b_i}$
are given in Arrhenius form
\begin{eqnarray}
 k_{f_i}=A^f_iT^{B^f_i} \exp\left(-C^f_i/T \right), \quad
 k_{b_i}=A^b_iT^{B^b_i} \exp\left(-C^b_i/T \right).
\end{eqnarray}
Since the above rate constants as given in Ref.~\cite{Evans81} are
based on ideal gas, the modification for a van der Waals gas as
suggested by Toegel \emph{et al.}~\cite{Tog02} is used.  As they
hinted, we can derive the equilibrium constant based on fugacity
for a van der Waals gas \cite{Reid}
\begin{eqnarray}
\label{Keq}
 K_{F_i} = \frac{ K_{P_i}^{\rm ideal}}{(1-
b\rho)^{\sum_{k=1}^N \nu_{ki}} \exp \left(-\frac{ \sum_{k=1}^N
\nu_{ki} b_k \rho_k} {1-b\rho} \right)} .
\end{eqnarray}
where $\nu_{ki}= \nu^{\prime\prime}_{ki}- \nu^\prime_{ki} $. Let
the corrected forward rate frozen $ k_{f_i}^\prime= k_{f_i} $. We
can obtain the corrected backward rate $ k_{b_i}^\prime $ as
\begin{eqnarray}
 k_{b_i}^\prime =\frac{ k_{f_i}^\prime }{ K_{P_i}^{\rm ideal} }=
\frac{ k_{b_i}}{(1- b\rho)^{^{\sum_{k=1}^N \nu_{ki}}}
\exp \left(- \frac{ \sum_{k=1}^N \nu_{ki} b_k \rho_k}
{1-b\rho} \right) }.
\end{eqnarray}
We will use $ k_{f_i}^\prime  $ and $ k_{b_i}^\prime $ in
Eq.~(\ref{progress_eq}). The Effects of the above modification
suggested in Ref.~\cite{Tog02} were found to suppress water vapor
dissociation to some extent when compared with the raw rate
constants $ k_{f_i}$ and $k_{b_i}$.

When involving the third-body reaction Eq.~(\ref{progress_eq})
becomes
\begin{equation}
q_i=\left[ \sum_{k=1}^N Z_{k,i}
\left(\frac{\rho_k}{M_k}\right)\right] \left[ k_{f_i}^\prime
\prod^N_{k=1}  \left(\frac{\rho_k}{M_k}\right)^{\nu^\prime_{ki}}
-k_{b_i}^\prime\prod^N_{k=1}\left(\frac{\rho_k}{M_k}\right)^{\nu^{\prime\prime}_{ki}}
\right],
\end{equation}
where $ Z_{k,i}$ is the third-body enhanced coefficient. Due to
length limitation, the detailed formulas for net production rates
and rate constants of collisional ionization, recombination, and
three-body recombination are not given here. One can refer to
Ref.~\cite{Ho02, Xu99} for detail.

\subsubsection{Mass and heat exchange at the bubble wall}

The evaporation-condensation process and diffusion of the noble
gas into the surrounding liquid  are included. The net evaporation
rate (mass per unit area and unit time) at the bubble wall is
given as follows \cite{Fuji80,Yasui97}:
\begin{eqnarray}
\label{eva_eq} \dot{m}_e= \frac{\alpha_M}{\sqrt{2\pi R_v}}
\left(\frac{P_{\rm sat} (T_{l, \rm int}) }{\sqrt {T_{l,\rm int}}
}- \frac{\Gamma P_{v, \rm int}}{\sqrt {T_{v, \rm int}} } \right),
\end{eqnarray}
where $\alpha_M$ is the accommodation coefficient (evaluated using
the formula in Ref.~\cite{Yasui97}) that shows which portion of
water vapor molecules hitting the liquid surface is absorbed by
this interface, $R_v$ is the gas constant of water vapor, $P_{\rm
sat}$ is the saturation vapor pressure at liquid temperature
$T_{l,\rm int}$, $P_{v,\rm int}$ is the partial pressure of water
vapor, and $\Gamma$ is the correction factor:
\begin{eqnarray}
\Gamma&=&\exp(-\Omega^2) -\Omega\sqrt{\pi} \left(1-\frac{2}{\sqrt{\pi}}
\int_0^\Omega \exp (-x^2)  dx\right), \\
\Omega &=& \frac{ \dot{m}_e} { P_{v,\rm int}} \sqrt{ \frac{R_v
T_{v,\rm int}} {2}} .
\end{eqnarray}
In this study, the jump of temperature across the interface is
assumed zero, thus $T_{l,\rm int}=T_{v, \rm int}$. Although
Eq.~(\ref{eva_eq}) is valid only below a critical point (for water
$T_{\text{cr}}\approx 647$ K), it is used throughout the whole
acoustic cycle for simplicity. The rate of mass diffusion of the
noble gas dissolved in the liquid  at the wall is
\begin{eqnarray}
\dot{m}_d= 4\pi R^2 D_l \left. \frac{\partial c}{\partial
r}\right|_{R},
\end{eqnarray}
where $R$ is the bubble radius, $D_l$ is the diffusion coefficient
of the noble gas, and $c$ is the mass concentration of the noble
gas dissolved in the liquid.

The boundary conditions of species and energy at the bubble
surface are derived by balancing the flux and source/sink of an
interface control volume with infinitesimal thickness. In doing
so, we assume that there are no uptakes of radical species or
surface reactions. The resulting boundary condition for gas
species is
\begin{eqnarray}
-J_i|_R + \dot{m}_{\text{tot}} Y_i = \dot{m}_e f^e_i +\dot{m}_d
f^d_i, \quad  i=1, \cdots   N,
\end{eqnarray}
where
\begin{eqnarray}
\dot{m}_{\text{tot}} = \dot{m}_e +\dot{m}_d,
\quad f^e_i=\left\{ \begin{array}{ll} 1,~ & i=\text{water vapor} \\
 0,~  & i=\text{other species}\end{array}\right.,
\quad f^d_i=\left\{ \begin{array}{ll} 1,~ & i=\text{noble gas} \\
 0,~ & i=\text{other species}\end{array}\right..
\end{eqnarray}
The boundary condition of energy is
\begin{eqnarray}
\label{heatflux_eq} \left[
 \lambda  \frac{\partial T}{\partial
r} -\sum_{i=1}^N \left(J_i-  \dot{m}_{\text{tot}} Y_i\right ) h_i \right]_R
 + \dot{m}_e L -\dot{m}_d \Delta H = \lambda_l \left. \frac{\partial T_l}{\partial r}\right|_R,
\end{eqnarray}
where $T_l$ and $\lambda_l$ are the temperature and thermal
conductivity of the liquid, respectively, $L$ is the latent heat
of evaporation of the liquid, and $\Delta H$ is the heat of
solution of noble gas into the liquid \cite{Reid,Yasui96}.

The gas and liquid velocities at the bubble surface and the
velocity of the bubble wall differ due to mass transfer. The
boundary condition for gas and liquid velocities are
\begin{eqnarray}
u\left.\right|_R=\dot{R}- \frac{\dot{m}_{\text{tot}}}{\rho},\quad
u_l\left.\right|_R=\dot{R}- \frac{\dot{m}_{\text{tot}}}{\rho_l}.
\end{eqnarray}
The mass and heat transfer boundary conditions are nonlinear. All
the boundary conditions for the gas dynamics have to be coupled
with the following  motion, temperature, and noble gas
concentration equations in the surrounding liquid.

\subsection{Motion, heat, and mass transport in the liquid}

The liquid flow outside the spherical bubble is accounted for with
different approximations for motion and heat (or mass) transport,
respectively. On the one hand, the Euler equations for the liquid
motion can be reduced to the ordinary differential equation for
the bubble radius known as the RP equation. The RP equation is
coupled with the NS equations through stress equilibrium condition
at the bubble wall. On the other hand, we assume that the fluid is
incompressible when accounting for heat and mass transfer in the
liquid. The separate treatment reduces the complexity of solving a
fully coupled hydrodynamic equations at sacrifice that shock waves
in the liquid can not be simulated well.

Because the mass transfer at the bubble wall results in very small
liquid velocity whose effect on the RP equation  can be ignored,
we let $u_l|_R=\dot{R}$. Thus we can use a form of the RP equation
\cite{Kel80, Pros86} that includes first order terms in the Mach
number $M=\dot{R}/C_{lb}$ and allows for variable speed of sound
in the water \cite{Kama87, Yuan98}:
\begin{equation}
\left(1-M \right)R \ddot{R}+ \frac 32 \left(1-
\frac 13 M \right) \dot{R}^2  = (1+M) \left[H_b-
\frac{1}{\rho_{l\infty}} P_s\left(t+\frac{R}{C_{l\infty}}\right)
\right] + \frac{R}{C_{lb}} \dot{H_b}\label{RP3_eq}  .
\end{equation}
Here subscripts $b$ and $\infty$ denote bubble wall and infinity,
respectively, $P_s(t)=-P_a \sin(2\pi f  t)$ is  the pressure of
the sound field with frequency $f$ and amplitude $P_a$. For water,
an equation of state of the modified Tait form
\begin{equation}
 \frac{P+B}{P_\infty+B}
=\left(\frac{\rho_l}{\rho_{l\infty}}\right)^n \ \  \label{tait}
\end{equation}
is used with  $B=3049.13$~bar and $n=7.15$. The enthalpy $H_b$ and
the speed of sound $C_{lb}$ of the liquid at the bubble surface
 are given by
\begin{eqnarray}
\label{enthalpy1} H_b &=& \int_{P_\infty}^{P_l}\frac{dP}{\rho_l} =
\frac{n}{n-1}\left(\frac{P_{lb}+B}{\rho_{lb}}
                        -\frac{P_\infty+B}{\rho_\infty}\right),\\
C^2_{lb}&=&\left.\frac{dP}{d\rho_l}\right|_b =  \frac{n(P_{lb}+B)}{\rho_{lb}} .
\end{eqnarray}
The  pressure $P_{lb}$ on the liquid side of the bubble surface is
related to the pressure $P(R,t)$ on the gas side of the bubble
surface by normal stress equilibrium condition
\begin{equation}
P(R,t)-\tau_{rr}|_{r=R}=P_{lb}+\frac{4\eta \dot{R}}{R}
+\frac{2\sigma}{R} \ \ ,
\end{equation}
where $\eta$ is the dynamic viscosity and $\sigma$ is the surface
tension. Their values depend on $T_l$, as formulated in the
Appendix of Refs.~\cite{Yasui96,Akha01}, respectively.

Both heat and mass transfer are taken into account although the
former is found to be more important to the bubble dynamics. The
equations for the water temperature and the mass concentration of
dissolved noble gas take a similar form:
\begin{eqnarray}
\label{diff_T_eq} \frac{\partial T_l}{\partial
t}+u_l\frac{\partial T_l}{\partial r} &= &\frac{ \lambda_l
}{\rho_l C_{P_l}}\frac{
\partial}{r^2\partial r}\left(r^2\frac{\partial T_l}{\partial r} \right)
,  \\
\label{diff_c_eq} \frac{\partial c}{\partial t}+u_l\frac{\partial
c}{\partial r} &=& D_l \frac{
\partial}{r^2\partial r}\left(r^2\frac{\partial c}{\partial r} \right),
\end{eqnarray}
where $C_{P_l}$ is the specific heat at constant pressure of the
liquid.  In Eqs.~(\ref{diff_T_eq}) and (\ref{diff_c_eq}), the
liquid velocity can be determined by the incompressible assumption
\begin{eqnarray}
u_l=\frac{\dot{R} R^2}{r^2}
\label{liquidv_eq}.
\end{eqnarray}
The boundary condition for the water
temperature are the continuity of heat flux
Eq.~(\ref{heatflux_eq}) and $T_l\left.\right|_{r=\infty}=
T_\infty$, and the boundary condition for the mass concentration is
\begin{eqnarray}
c\left.\right|_{r=R}=\frac{c_0(T_\infty,P_0)}{P_0}
P_{\text{no}}(R,t), \quad  c\left.\right|_{r=\infty}= c_\infty,
\end{eqnarray}
where $c_0$ is the saturated dissolved gas concentration at
$T_\infty$ and $P_0$, and $P_{\text{no}}$ is the partial pressure
of the noble gas on the internal side of the bubble interface.

\subsection{Numerical method}

To exploit the advantage of a stationary Eulerian meshes, we use $
x = r/R(t)$ to transform the NS equations (\ref{NS_eq}) to a form
in fixed domain $x\in [0,1]$ as done in our earlier work
\cite{Yuan98}. The transport equations for water temperature $T_l$
and mass concentration $c$ in domain $r\in [R,\infty ] $ are
transformed into diffusion-type equations in domain $z\in [0,1]$
through two consecutive coordinate transformations
\cite{Plesz,Gros77} with the aid of Eq.~(\ref{liquidv_eq}). The
details were given in Refs.~\cite{Hilg96, Yuan98}.

We apply the second-order upwind total-variational-diminishing
(TVD) scheme \cite{Yee} to the inviscid flux terms and the central
difference to the diffusive terms of the NS equations. The
temporal discretization differs from Ref.~\cite{Yuan98} in that we
now use an Adams-Bashforth explicit scheme for the convective and
spherical coordinate terms, and Crank-Nicolson implicit scheme for
diffusive and chemical source terms. The implicit treatment is to
overcome the stiffness problem due to diffusive transport and
chemical source terms. The trapezoidal rule and central difference
are used for the water temperature and concentration equations. A
predictor-corrector method is used for the RP equation. We use 400
grid points for the NS equations and 100 points for the water
temperature and gas concentration equations.

\section{Optical power radiated by the bubble}
\label{optic}

We compute SBSL based on the weakly ionized gas model of
Hilgenfeldt \emph{et al.}~\cite{HilgPHF99}, which was most
thoroughly studied and remarkably successful \cite{Ham01}. In more
general cases with nonuniform bubble interior, the bubble has an
optically thin radiating/absorbing outer shell, and may have a
blackbody inner core when the opacity is large enough
\cite{Moss97,Moss99}. We previously applied this generic version
to a pure argon bubble \cite{Ho01,Ho02}. In this paper, we apply
the formulas of photon absorption coefficients
\cite{HilgPHF99,Zeld66} to the gas mixture in the bubble. The
overall photon absorption coefficient $\kappa_{\lambda}^{\rm tot}$
is the sum of contributions from all species
\begin{equation}
  \kappa_{\lambda}^{\rm tot}=\sum_{i=1}^N \kappa_{\lambda,i}
= \sum_{i=1}^N \left( \kappa_{\lambda,i}^{\rm ff+} +
\kappa_{\lambda,i}^{\rm
 ff0} + \kappa_{\lambda,i}^{\rm bf} \right),
\end{equation}
where $\kappa_{\lambda,i}^{\rm ff+}$ is the absorption due to the
free-free interaction of electron and ions,
$\kappa_{\lambda,i}^{\rm ff0}$ is the  absorption due to free-free
interactions of electrons and neutral atoms, and
$\kappa_{\lambda,i}^{\rm bf}$ is the absorption by bound-free
ionization of already excited atoms. The ionization potentials
used can be found in Ref.~\cite{Brown}. The bound-bound absorption
is not accounted for and the modification to electron-neutral-atom
bremsstrahlung \cite{Fromm98,Ham00} is not adopted for the time
being.

The finite opacity model given in Refs.~\cite{Ho01,Ho02} (without
the $\Theta$ correction) is used to compute the total spectral
radiance (power emitted per wavelength interval) of the bubble
content at wavelength $\lambda$
\begin{eqnarray}
\label{thinp_eq}
 P_{\lambda}^{\rm Pl}(t)
&=&\int_{R_c}^R 16\pi \kappa_\lambda^{\rm
tot}(r,t)R_{\lambda}^{\rm Pl}(r,t)
 \exp\left(-\int_r^R\kappa_\lambda^{\rm tot}(r',t)dr'\right)
   r^2dr \\  \nonumber
& &+ 4\pi{R^2_c}R_{\lambda}^{\rm Pl}(R_c,t)
 \exp\left(-\int_{R_c}^R\kappa_\lambda^{\rm tot}(r,t)dr\right)
  \;\;,
\end{eqnarray}
where
\begin{equation}
 R_{\lambda}^{\rm Pl}[ T(r,t) ]
=\frac{2\pi hc^2}{\lambda^5} \frac{1}{\exp(hc/\lambda k_BT)-1}
\end{equation}
is the spectral emissive power (energy per unit time, wavelength
interval, and projected surface area) with the Planck and
Boltzmann constants $h$ and $k_B$, and the light speed in vacuum
$c$. The time-dependent radius of the blackbody core $R_c$ can be
defined by the expression \cite{Moss99}
\begin{equation}
 \int_{R_c}^R \bar{\kappa}^{\rm tot} (r,t) dr
=1, \label{epc_eq}
\end{equation}
where $\bar{\kappa}^{\rm tot} $ is the wavelength-averaged
absorption coefficient between 200 and 800 nm.
Equation~(\ref{epc_eq}) implies that if radiation from a spherical
surface at depth  $R_c$ is damped to some extent (the optical
depth being 1), then radiation from further interior is completely
opaque to an outside observer. The spherical surface $R_c$ serves
as the surface of a blackbody in place of radiations from the
interior. The determination of $R_c$ starts from the outermost. If
$\bar{\kappa}^{\rm tot}$ is sufficiently large, there will be a
finite-size blackbody core $ 0 < R_c \leq R $ such that
Eq.~(\ref{epc_eq}) is satisfied; if $\bar{\kappa}^{\rm tot}$ is
small, the left hand side of Eq.~(\ref{epc_eq})  will be less than
1 even if $R_c$=0, implying that the bubble is optically thin. The
calculated photon absorption coefficients indicate that the bubble
is always optically thin, $R_c\equiv 0$. However, in order to see
how a finite-size blackbody model behaves, we intentionally
amplify $\bar{\kappa}^{\rm tot}$ by a free parameter $E_c $ so
that
\begin{equation}
 \int_{R_c}^R E_c \cdot  \bar{\kappa}^{\rm tot} (r,t) dr
=1 \label{amp}
\end{equation}
will give a nonzero $R_c$ during the collapse stage. It is evident
that larger $E_c$ makes $R_c$ closer to $R$. With $R_c(t)$ at
hand, we can calculate the light emission by Eq.~(\ref{thinp_eq})
together with original $\kappa_\lambda^{\rm tot}$, whose small
quantity makes the second term  in Eq.~(\ref{thinp_eq}) dominant.
For convenience of discussion we denote $R_c\equiv 0$ as the
optically thin model and $R_c
>0$ as the finite-size blackbody model. It is remarked that
the finite-size blackbody model is physical if $E_c=1 $, and is
\emph{ad hoc} if $E_c >1$. For fitting purpose, $E_c$ is different
from case to case but remains fixed during an acoustic cycle.

It is meaningful to look at the light pulses and spectra. The
integration of the spectral radiance over a suitable wavelength
intervals ($\lambda_\text{UV}=200~\text{nm}\sim\lambda_r= 800
~\text{nm}$) gives the total power emitted into the measurable
part of the spectrum, and integration over one acoustic period
$T_s$ gives the light spectrum that can be compared with the
experimental results
\begin{equation}
 P^{\rm Pl}(t) = \int^{\lambda_r}_{\lambda_{\rm UV}}  P_{\lambda}^{\rm Pl}
(t)  d\lambda, \quad
 S^{\rm Pl}_\lambda =\frac{1}{T_s} \int^{T_s}_0  P_{\lambda}^{\rm Pl} (t)dt
\end{equation}

\section{ NUMERICAL RESULTS}

It is well known that the ambient bubble radius $R_0$ depends on
experimentally controllable parameters such as the driving
pressure amplitude $P_a$, the water temperature $T_\infty$, and
the gas concentration dissolved in the water $c_\infty$.  A
problem with past SL spectrum measurements was they seldom gave
the key parameters $P_a$ and $R_0$ at the same time. This left
freedom for theoretical studies to fit experimental data using
different $P_a$ and $R_0$. The present study tries to use the same
parameters as those in previous literatures. For comparison with
other calculations, the parameters used here are $P_a =1.2$~bar,
$T_\infty=298$~K (Storey \cite{Stor00}) and $P_a =1.35$~atm,
$T_\infty=300$~K (Xu \cite{Xu03}) for identical equilibrium radius
$R_0 = 4.5$~$\mu$m and driving frequency $f=26.5$~kHz, and
$P_a=1.4$~bar for $R_0 = 6.0$~$\mu$m, $f=20.6$~kHz,
$T_\infty=293.15 $~K ( Moss \cite{Moss99}). The dissolved gas
concentration is $c_\infty/c_0=P_{\rm{no}_{\infty}}/P_0=0.395\%$
(3 Torr partial pressure) for all above cases.  For comparison
with the experiment \cite{Vaz01}, $R_0=4.5~\mu$m~(He),
$5.5~\mu$m~(Xe), $f=42$~kHz, $T_\infty=296.15$~K,
$P_{\rm{no}_{\infty}}=$ 150 Torr (He) and  3 Torr (Xe), while
$P_a$ is adjustable. Other parameters are  $P_\infty=101325\,{\rm
Pa}$, $\rho_{l\infty} =996.6\,{\rm kg\,m}^{-3}$, $k_l=0.609\,{\rm
W\,m}^{-1}\,{\rm K}^{-1}$,  $C_{{\rm P}l}=4179\,{\rm
J\,kg}^{-1}\,{\rm K}^{-1}$,  and $D_l=2 \times 10^{-9}~ {\rm m}^2
/$s. The van der Waals excluded volumes are given in Table I.
Initial number densities of ions and electrons are estimated using
the Saha equation \cite{Zeld66}. The initial bubble content
contains 2\% molar fraction water vapor. This number seems
arbitrary, but our results are based on the second acoustic cycle
when initial disturbances are presumed to be decayed.
\begin{table*}[hbt]
\caption{van der Waals excluded volumes.}
\begin{ruledtabular}
\begin{tabular}{c c c c c c c c c c c c c }
Species & He \& ions & Ar \& ions & Xe \& ions & H$_2$O & OH & H &
H$^+$ & O & O$^+$ & H$_2$ & O$_2$ & e$^-$
\\ \hline
$\tilde{b}_i$ (cm$^3$/mol)  & 23.7  & 32.2 & 51.0 & 30.5 & 15.25
  & 4.98& 36.8 & 2.77 & 27.7 & 26.6 & 31.8 &  0.0
\end{tabular}
\end{ruledtabular}
\end{table*}

\subsection{Effects of chemical reactions}

Figure 1 shows one forcing period of the radius of an argon bubble
corresponding to case I (pure noble gas in the bubble), case II
(with water vapor but no chemical reactions), and case III (with
water vapor and chemical reactions)  as labeled in
Ref.~\cite{Stor00}. One can see that the difference between II and
III is almost indiscernible, but that between I and the latter two
is large. The existence of water vapor increases the maximum
radius and delays the collapse. In spite of little difference
between II and III in the $R$-$t$ curve, large difference occurs
for thermodynamic variables at collapse. Table II shows comparison
of some quantities. The maximum radius of the present calculation
is smaller than that of Story \cite{Stor00}. This is mainly
because the acoustic forcing terms in Eq.~(\ref{RP3_eq}) is
separate from $H_b$ rather than absorbed in $P_\infty$ as treated
in Refs.~\cite{Stor99,Stor00} that magnified $P_a$ by a factor of
$n/(n-1)$ [see Eq.~(\ref{enthalpy1}) ]. Since there are much
differences between the present model and Storey's, quantitative
discrepancies are expectable for the extreme values. Both models
predicted temperature reduction from case I to III. However, the
present result show that temperature is slightly reduced from I to
II, but heavily from II to III. The slight reduction is due to the
compensating effects of increased compression ratio
($R_\text{max}/R_\text{min}$) and reduced ratio of specific heats
\cite{Stor00}.  The larger reduction from 17 000~K (case II) to
8900~K (case III) indicates the significant effect of chemical
reactions in reducing the temperature.

\begin{table}[htp]
\caption{Comparison of extreme values for an argon bubble
$R_0=4.5~\mu$m, $P_a=1.2$~bar. The amount of water vapor in II is
in mole fraction and evaluated at the moment of $R_\text{min}$.}
\begin{ruledtabular}
\begin{tabular}{c c c c c c c }
&\multicolumn{2}{c}{ $R_\text{max}/R_\text{min} (\mu\text{m})$ } &
\multicolumn{2}{c}{ $T_\text{max}$~(K) }& \multicolumn{2}{c}{total vapor (\%)}\\
case & present & Ref.~\cite{Stor00}&  present &  \cite{Stor00}
& present &  \cite{Stor00} \\
\hline
I & 25.4/0.88   & 28.0/0.80 &  17900 &20900 & &  \\
II & 28.7/0.76   & 31.3/0.70  &17000  & 9700 & 7.7   &
14\footnote{Reference~\cite{Stor00} did not specify the exact
moment
for this value.}   \\
III & 28.9/0.70   & 31.7/0.65 & 8900 & 7000 & &
\end{tabular}
\end{ruledtabular}
\end{table}

The effects of chemical reactions on thermodynamic variables are
best reflected in the distributions of temperature  and chemical
yields inside the bubble. Figure~2 shows snapshots of the spatial
profiles of temperature around the moment of minimum bubble
radius. It can be seen that temperatures are considerably reduced
in the reacting case III. Figure~3 shows the numbers of molecules
of different species and temperature at the bubble center as a
function of time. Note that in the first acoustic cycle the water
vapor begins to dissociate appreciably at $t=-2\sim -1$~ns, while
in the beginning of the second cycle there are already some
amounts of H$_2$ and O$_2$ gases accumulated. The chemical
reactions occur in a time scale of several nanoseconds, producing
considerable amounts of H, O, OH radicals and H$_2$ and O$_2$
gases. It is remarked that the prediction of chemical products is
very difficult as the reaction mechanisms and phase change
processes are largely unknown under extreme conditions in a SL
bubble.

Next we compare a $R_0=6.0~\mu$m Ar bubble driven at
$P_a=1.4$~bar, which was labeled as A1 in Ref.~\cite{Moss99}. The
present temperatures at the bubble center are respectively 109
600~K, 34~700~K, and 16~200~K for cases I$-$III in Fig.~1,
suggesting that both the reduced ratio of specific heats due to
the presence of water vapor and the chemical reactions contribute
significantly to the reduction of temperatures. The amount of
trapped water vapor at the moment of $R_\text{min}$ for case II
occupies 23\% molar fraction, smaller than 33 \% \cite{Stor00}.
Figure~4 shows several snapshots of the spatial profiles of
thermodynamic variables. A main feature is that only compression
waves occur. As seen from the velocity profile, a compression wave
moves outward at $t_4$ and $t_5$, reflects from the bubble wall
and moves inward at $t_6$.  This result is different from that of
Moss \emph{et al.} \cite{Moss99}, where shock waves were reported.
A possible reason is that the formation of a shock is sensitive to
differences in equations of state, accommodation coefficients,
chemical reactions, and treatments of the liquid motion, and so
on. Another feature in Fig.~4 is that the temperature in the inner
zone is reduced more severely than in the outer zone at $t_1,t_2,
t_7$, and $t_8$ when the compression wave is not strong. This is
because considerable water vapor is trapped in the inner zone of
an Ar bubble as a result of thermal diffusion \cite{Stor99, Xu03}
(which states that a heavier species tends to diffuse toward the
cooler region) and dissociates there, thus peak temperature is not
at the center, but at some place close to the bubble interface.
For a lighter-than-water-vapor He bubble, less water vapor is
congregated in the central zone, and temperature peak will be
located at the bubble center as [will be shown in Fig.~6 (b)].
Figure~5 shows one snapshot of the number density distributions
and the degrees of ionization. It can be seen that the amounts of
products due to chemical reactions are considerable, while the
degrees of ionization are quite small (the maximum being 2.8\% for
O$^+$), contrary to significant ionizations when water vapor was
not taken into account \cite{Ho02}. The degrees of ionization of H
and O far exceed that of Ar although Ar atom is more ample in
quantity.

The numerical results in this subsection demonstrates that the
trapped water vapor and the ensuing endothermic chemical reactions
significantly reduce the temperature, resulting in very low
degrees of ionization. The chemical reactions can produce
considerable amounts of atomic and molecular species, some of
which, such as H and O atoms, are easier to ionize than a noble
gas atom such as He or Ar. As the evaporation is a robust process
in the bubble oscillation, chemical radicals will have significant
influence on SBSL mechanism as will be shown in Sec.~IV~C.

\subsection {Effects of noble gas types}

Previous numerical studies pointed that shock formation depends
sensitively on, among other factors \cite{Yuan98}, the amount of
water vapor \cite{Moss99, Xu03} and its distribution
\cite{Stor00}. Xu {\it et al.} \cite{Xu03} showed that shock waves
develop in a bubble filled with 70\% Xe and 30\% (mole fractions)
water vapor, but no shocks occur for similarly filled Ar or He
bubbles. With evaporation-condensation process and chemical
reactions taken into account, we are able to investigate effects
of noble gases on the thermodynamic processes more realistically.
We calculated  Xe, Ar and He bubbles using the same $R_0, P_a, f$,
and $T_\infty$ as those in Ref.~\cite{Xu03}.

Figures~6(a) and 6(b) show snapshots of the spatial profiles of
thermodynamic variables around the moment of minimum bubble radius
for Xe and He bubbles. Snapshots of the Ar bubble are similar to
those in Fig.~4, thus are not shown here. In the Xe bubble
[Fig.~6(a)], it is seen that an inward-going compression wave at
$t_2$ evolves into a strong outward-going shock at $t_3$. The
first focusing of the shock happened between $t_2$ and $t_3$ leads
to extreme high temperatures ($>10^6$~K), but the duration is very
short ($<$ 1 ps) and the region is confined to the center ($r<
0.005~\mu$m). However, Fig.~6(b) shows that only weak compression
waves occur in the He bubble. These results are in qualitative
agreement with those of Xu \emph{et al.}~\cite{Xu03}. Note that
the temperature peaks are often at the center in He bubble except
when a wavy disturbance reflects from the bubble wall at $t_4$.
This feature
 mainly results from the thermal diffusion between the light He gas
and the heavy water vapor as mentioned in previous subsection.

 Figure~7(a) and 7(b) show one snapshot of the
number density and the compositional distributions at the moment
of minimum bubble radius for Xe and He bubbles. In both bubbles
there are significant numbers of chemical products, especially H,
O, OH, and H$_2$. However, the right figure in Fig.~7(a) indicates
that atomic species in Xe bubble are significantly ionized only in
the central zone ($r < 0.07 R$), while the right figure in
Fig.~7(b) shows that the degrees of ionization in He bubble are
small, especially that for He gas. The situation for Ar bubble is
found to be in between [Fig.~5(b)]. This suggests that chemical
products of the water vapor may play  an increasingly important
role in SBSL from Xe, Ar, to He bubbles.

\subsection {Calculated light spectra and pulses}
\label{sect_light}

We  shall compute the emitted lights by using the optically thin
model and the \emph{ad hoc} finite-size blackbody model. We fit
our calculations to Fig.~2 of a recalibrated experiment
\cite{Vaz01} under the same parameters as given in the beginning
of this section. The former model has only $P_a$ while the latter
has $P_a$ and $E_c$ in Eq.~(\ref{amp}) as fitting parameters.

Figure~8 shows comparison of the spectral radiances.  It is seen
that the optically thin model does not match well with the
experimental spectrum of Xe bubble, but the finite-size blackbody
model matches well. The fitting $P_a=1.55~$atm of the optically
thin model seems to be out of the stable SBSL range ($1.2-1.5$
atm), while the fitting $P_a=1.28~$atm of the \emph{ad hoc}
finite-size blackbody is within the range, of course with the help
of a large value $E_c=1.8\times 10^4$. (Fitting using
$P_a=1.49~$atm and  a smaller value, $E_c=60$  can also give
similar spectrum but the resulting FWHM is only 40 ps, much
shorter than experimental 200 ps.) However, it can be seen that
either models are unable to fit the spectrum of He bubble. The
finite-size blackbody is much better than the optically thin model
as the latter deviates severely from the experiment. The maximum
temperatures at $P_a=1.28~$atm for Xe bubble and at $P_a=1.45~$atm
for He bubble are 8600 and 16 700~K, respectively, which are
comparable to the blackbody fitting temperatures 8000 (Xe) and 20
400~K (He) used in Ref.~\cite{Vaz01}.

Figure~9 compares the time variations of the normalized power for
the ``measurable," ``UV'' (300$-$400 nm), and ``red'' (590$-$650
nm) wavelength intervals \cite{Gom}. Both models show good
wave-length independence of light pulse, a key ingredient of SBSL
thought by several researchers \cite{Gom, HilgPHF99}.  One curves
``A," the optically thin model gives  FWHM of 26 ps for Xe bubble,
and 13 ps for  He bubble, which are much shorter than the
experimental flash widths of 200 ps (Xe) and 100 ps (He)
\cite{Vaz01}. On curves ``B," the finite-size blackbody model
predicts  FWHM of 185 ps for Xe bubble, and 18 ps for He bubble,
better than the optically thin model. Figure~10 shows the
variations of the blackbody core and bubble radius with time. One
can see that the blackbody core appears abruptly, attains maximum
around the moment of minimum bubble radius, and disappears
suddenly. The short life of the blackbody core explains why this
blackbody model also shows wave-length independence of the light
pulse similar to the optically thin model: The quick rise and fall
of $R_c$ in accordance with the variation of the photon absorption
coefficients cut down the long fall time for red light
\cite{HilgPHF99}.

Figure~11 shows the visible light powers contributed from the
total and partial species as computed by the optically thin model.
In Fig.~11(a) we see that the power from xenon is dominant, while
that from the water vapor and its chemical products contributes a
little. But in Fig.~11(b), the power from helium is small, while
the water vapor and its chemical products contribute dominantly
with H and O radicals being the primary ones. Although the
absolute values of the light powers from the optically thin model
are quite small, the relative contributions to the total power
verify previous postulation that light emission from radicals
generated from water vapor dissociation may dominate SBSL in the
He bubble \cite{HilgPHF99}.

The failure of the optically thin model to match with the
experiment and the improvement by adding an \emph{ad hoc}
finite-size blackbody core to it raise contradiction. The major
reason for the failure of the optically thin model is the
significant reduction of temperature due to the existence of water
vapor and endothermic chemical reactions. We remark that the
present \emph{ad hoc} blackbody is not the unique way for better
fitting, but is shown better than other \emph{ad hoc} ways such as
multiplying the photon absorption coefficient by an arbitrary
factor. We did not exclude the possibility that the optically thin
model could fit well to the experimental data should higher
temperature or increased opacity be obtained in whatever a natural
way. As a numerical study, we have made tests of model sensitivity
to parameters. The evaporation-condensation process and chemical
reaction rates are important for the modeled temperature. We used
the accommodation coefficient $\alpha_M=0.4 $ \cite{Stor00} as
well as the formula \cite{Yasui97}, and found that the former led
to higher temperature so that we could use $P_a=1.7~$atm to obtain
the same spectrum of He bubble as the one obtained by using
$P_a=2.0~$atm in Fig.~8. However, the results were still unable to
fit to the data. As to the chemical reaction rates, we compared
two different reaction rate sets \cite {Yasui97,Evans81}. There
was slight difference in quantity and what we presented here was
the one corresponding to higher temperature \cite{Evans81}. We
have already adopted the modified chemical equilibrium constant
(\ref{Keq}) for a van der Waals gas. This modification was claimed
to let the optically thin model give sufficient light emission
\cite{Tog02}, but we found  while it raised temperature to some
extent, it could not result in enough light emission. Therefore,
it will be better to pursue other ways to reconcile the conflict.
One conjecture was if there existed a mechanism that would greatly
increase the photon absorption coefficient of the highly
compressed bubble content \cite{Bren02, Vaz01}, such as the
lowering of ionization potentials \cite{Ham01}. it is also wished
to have better theories to compute the photon absorption
coefficients of a very dense gas mixture, and to take into account
other light emission processes due to the existence of chemical
products. Meanwhile further efforts are necessary to reduce
modelling uncertainties such as the chemical reaction rates under
high-pressure and high-temperature conditions, and the description
of the surrounding liquid motion. The RP equation approach in this
and other studies \cite{Stor00} should be more critically compared
with the full hydrodynamic equation approach \cite{Moss99,
Burn01}.

\section {Conclusions}

A refined hydrochemical model is presented to simulate the complex
processes inside a sonoluminescing bubble. The numerical
simulations of Xe, Ar, and He bubbles indicate that the trapped
water vapor and its endothermic reactions reduce the temperature
significantly. In the stable SBSL range, at most compression waves
can appear in He or Ar bubbles, while shock waves can occur in Xe
bubbles only for higher driving amplitudes. The lower temperature
in the bubble rarely leads to appreciable ionization except for Xe
bubble at the center during the shock wave focusing. The chemical
radicals generated from water vapor dissociations become
increasingly important in the light emission from Xe, Ar, to He
bubbles. Particularly, H and O radicals are shown  to be the
primary light-emitting matters in He bubbles.

The key finding of this study is that the optically thin thermal
emission model was unable to match with experimental data mainly
due to the reduced temperatures in the bubble. The introduction of
a finite-size blackbody core made the calculated light spectra and
pulse widths match better with experimental ones. The present
expertise to define an optically thick region is \emph{ad hoc} and
only serves to illustrate one possible improving direction.
Improvements by physical considerations such as the lowering of
ionization potentials and the refinement of the photon absorption
processes under the extreme conditions of sonoluminescence, and by
reduction of modeling uncertainties, are worthy of further
investigation.

\begin{acknowledgments}
The author thanks M.-C.~Chu and Yu An for meaningful discussions.
This work is supported by National Natural Science Foundation of
China (Grant Nos. G10172089, G10476032) and State Key Program of
Basic Research (Grant No. G1999032801).
\end{acknowledgments}


\begin{thebibliography}{99}

\let\ul=\underbar
\def\AP#1{{\it Ann.~Phys.~ }{ #1}}
\def\PRP#1{{\it Phys.~Rep.~ }{ #1}}
\def\APP#1{{\it Act.~Phys.~Pol.~ }{ #1}}
\def\PTP#1{{\it Prog.~Theor.~Phys.~ }{ #1}}
\def\PRD#1{{\it Phys.~Rev.~D}{ #1}}
\def\PRC#1{{\it Phys.~Rev.~C}{ #1}}
\def\PRL#1{{\it Phys.~Rev.~Lett.~}{#1}}
\def\PL#1{{\it Phys.~Lett.~ }{ #1}}
\def\NP#1{{\it Nucl.~Phys.~ }{ #1}}
\def\ZP#1{{\it Z.~Phys.~ }{ #1}}

\bibitem{Gait92}D. Gaitan, L. Crum, C. Church, and R. Roy,
J.~Acoust.~Soc.~Am.~{\bf 91}, 3166 (1992).

\bibitem{Barber97} B. Barber, R. Hiller,  R. L\"{o}fstedt,
S. Putterman, and K. Weninger,  Phys.~Rep.~{\bf 281}, 65 (1997).

\bibitem{Bren02}M. Brenner, S. Hilgenfeldt, and D. Lohse,
Rev.~Mod.~Phys., {\bf 74}, 425 (2002).

\bibitem{Gom}B. Gompf, R. G\"{u}nther, G. Nick, R. Pecha, and
W. Eisenmenger, Phys.~Rev.~Lett.~{\bf 79}, 1405 (1997).

\bibitem{Hil}R. Hiller, S. Putterman, and K. Weninger,
Phys.~Rev.~Lett.~{\bf 80}, 1090 (1998).

\bibitem{Mor}M. Moran and D. Sweider, Phys.~Rev.~Lett.~{\bf 80}, 4987 (1998).

\bibitem{Hilg_nature}S. Hilgenfeldt, S. Grossmann, and D. Lohse,
Nature (London)~{\bf 398}, 402 (1999).

\bibitem{HilgPHF99}S. Hilgenfeldt, S. Grossmann,  and D. Lohse,
Phys.~Fluids~{\bf 11}, 1318 (1999). We note that the term ``vacuum
permeability" denoted as $\epsilon_0$  in this reference as well
as in Ref.~\cite{Ho02} should be the ``vacuum permitivity"
consistent with \cite{Zeld66}.

\bibitem{Ham00}D. Hammer and L. Frommhold,  Phys.~Rev.~Lett.~{\bf 85},
1326 (2000).

\bibitem{Ham01}D. Hammer and L. Frommhold, J. Mod. Opt. \textbf{48}, 239 (2001).

\bibitem{Moss97}W. Moss, D. Clarke, and D. Young,
Science {\bf 276}, 1398 (1997).

\bibitem{Moss99}W. Moss,  D. Young, J. Harte,
J. Levatin, B. Rozsnyai, G. Zimmerman, and  I. Zimmerman,
Phys.~Rev.~E~{\bf 59}, 2986 (1999).

\bibitem{Burn01}P. Burnnet, D. Chambers, D. Heading, A. Machacek, M. Schnittker,  W. Moss,
P. Young, S. Rose, R. Lee, and J. Wark, J.~Phys.~B  \textbf{34},
L511 (2001).

\bibitem{Ho01}C. Y. Ho, L. Yuan, M.-C. Chu, P. T. Leung,  and
W. Wei, Europhys.~Lett. {\bf 56}, 891  (2001).

\bibitem{Ho02}C. Y. Ho, L. Yuan, M.-C. Chu, P. T. Leung,  and
W. Wei, Phys.~Rev.~E~{\bf 65}, 041201 (2002).

\bibitem{Vaz01} G. Vazquez, C. Camara, S. Putterman, and K. Weninger,
Opt.~Lett. \textbf{26}, 575 (2001).

\bibitem{Vuong96}V. Vuong  and A. Szeri,
Phys.~Fluids {\bf 8}, 2354 (1996).

\bibitem{Yuan98}L. Yuan, H. Y. Cheng, M.-C. Chu, and P. T. Leung,
Phys.~Rev.~E~{\bf 57}, 4265 (1998).

\bibitem{Cheng}H. Y.  Cheng, M.-C. Chu, P.T. Leung, and L. Yuan,
Phys.~Rev.~E~{\bf 58}, R2705 (1998).

\bibitem{Fuji80}S.  Fujikawa  and A. Akamatsu , J.~Fluid Mech.~\textbf{97},
481 (1980).

\bibitem{Kama93}V. Kamath,  A. Prosperetti, and F. Egolfopoulos,
J.~Acoust.~Soc.~Am.~{\bf94}, 248 (1993).

\bibitem{Soch97}S. Sochard, A. Wilhelm, and H. Delmas, Ultranson.~Sonochem.
\textbf{4}, 77 (1997).

\bibitem{Gon98}C. Gong  and D. Hart,
J.~Acoust.~Soc.~Am.~{\bf104}, 2675 (1998).

\bibitem{Yasui97}K. Yasui, Phys.~Rev.~E {\bf 56}, 6750 (1997).

\bibitem{Sus99}K. Suslick, Y. Didenko, M.F. Fang, T. Hyeon, K. Kolbeck, W. McNamara,
M. Mdleleni, and M. Wong, Philos.~Trans.~R.~Soc.~London, Ser.~A
\textbf{357}, 335 (1999).


\bibitem{Stor00}B. Storey and A. Szeri, Proc.~Roy.~Soc.~London, Ser. A {\bf 456},
1685 (2000).

\bibitem{Tog00}R. Toegel, B. Gompf, R. Pecha, and D. Lohse,
Phys.~Rev.~Lett.~{\bf 85}, 3165 (2000).

\bibitem{Tog02}R. Toegel, S. Hilgenfeldt, and D. Lohse,
Phys.~Rev.~Lett.~{\bf 88}, 034301 (2002).

\bibitem{Xie03}C. C . Xie, Y. An, and C. F. Ying, Acta Phys. Sin.  \textbf{52}, 102
(2003).

\bibitem{Akha01}I. Akhatov, O. Lindau, A. Topolnikov, R. Mettin,
N. Vakhitova, and W. Lauterborn, Phys.~Fluids~\textbf{13}, 2805
(2001).

\bibitem{Xu03}N.  Xu, R. Apfel, A. Khong, X. W.  Hu, and L. Wang,
Phys.~Rev.~E.~{\bf 68}, 016309 (2003).

\bibitem{Poist}T. Poinsot and D. Veynante, \emph{Theoretical and Numerical
Combustion}, (R.T. Edwards, Flourtown, PA, 2001).

\bibitem{Hirsc54} J. Hirschfelder, C. Curtiss,  and R. Bird,
\emph{Molecular theory of gases and liquids}  (Wiley, New York,
1954).

\bibitem{Reid}R. Reid, J. Prausnitz, and T. Sherwood,
\emph{The properties of gases and liquids}, 3rd ed. ( McGraw-Hill,
New York, 1977).

\bibitem{Stor99}B. Storey and A. Szeri, J.~Fluid Mech.~{\bf 396},
203 (1999).

\bibitem{Gupta}R. Gupta, J. Yos, and R. Thompson,
``A review of reaction rates and thermodynamic and transport
properties for the 11-species air model for chemical and thermal
nonequilibrium calculations to 30 000 K'', NASA TM 101528, 1989
(unpublished).

\bibitem{Wu93}C. C. Wu and P. Roberts, Phys.~Rev.~Lett.~{\bf 70},
3424 (1993).

\bibitem{GRI3}Smith G \emph{et al.}, GRI-Mech~3.0, 1999,
http://www.me.berkeley.edu/gri\_mech/

\bibitem{Evans81}J. Evans and C. Schexnayder,  AIAA~J.~\textbf{18},
188 (1980).

\bibitem{Xu99} N.  Xu, L. Wang,  and X. W. Hu,
Phys.~Rev.~Lett.~{\bf 83}, 2441 (1999).

\bibitem{Yasui96} K. Yasui, J.~Phys.~Soc.~Jpn. \textbf{65}, 2830 (1996).

\bibitem{Kel80}J. Keller and M. Miksis,
J.~Acoust.~Soc.~Am.~{\bf68}, 628 (1980).

\bibitem{Pros86}A.  Prosperetti and A. Lezzi,
J.~Fluid Mech.~{\bf 168}, 457 (1986).

\bibitem{Kama87}V. Kamath  and A. Prosperetti,
J.~Acoust.~Soc.~Am.~{\bf85}, 1538 (1987).

\bibitem{Plesz}M. Plesset and S. Zwick,
J.~Appl.~Phys.~{\bf 23}, 95 (1952).

\bibitem{Gros77}C. Grosh and S. Orszag,
J.~Comput.~Phys.~{\bf 25}, 273 (1977).

\bibitem{Hilg96}S. Hilgenfeldt, D. Lohse, and M. Brenner,
Phys.~Fluids~{\bf 8}, 2808 (1996).

\bibitem{Yee}H. C. Yee, ``A class of high-resolution explicit and implicit
shock-capturing methods'', NASA TM 101088, 1989 (unpublished).


\bibitem{Zeld66}Y. Zeldovich and Y. Raizer, \emph{Physics of Shock
Waves and High-temperature Hydrodynamic Phenomena}, edited by
W.~D.~Hayes and R.~F.~Probstein (Academic Press, New York, 1966),
Vol.1.

\bibitem{Brown}S. C.  Brown, \emph{Basic data of plasma physics} (MIT Press,
Cambridge,  1966).

\bibitem{Fromm98}L.  Frommhold, Phys.~Rev.~E~{\bf 58}, 1899 (1998).






\end{thebibliography}

\begin{figure}[hbp]
\begin{minipage}[t]{0.49\textwidth}
\includegraphics[width= 8.5cm ]{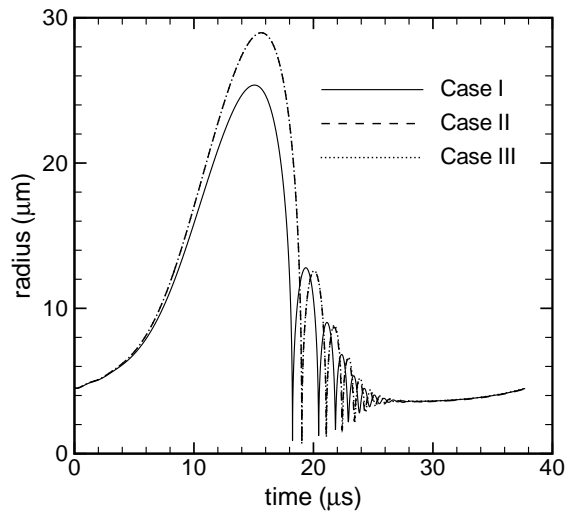}
\end{minipage}
\caption{Radius of an argon bubble vs  time over one acoustic
period for the same parameters as used in Ref.~\cite{Stor00}:
$R_0=4.5~\mu$m, $P_a=1.2~$bar, $f=26.5~$kHz, $T_\infty=298$~K.
Case I: without phase change and chemistry; case II: with phase
change but without chemistry; case III: with phase change and
chemistry. } \label{fig1}
\end{figure}

\begin{figure}[]
\begin{minipage}[t]{0.49\textwidth}
\includegraphics[width= 8cm ]{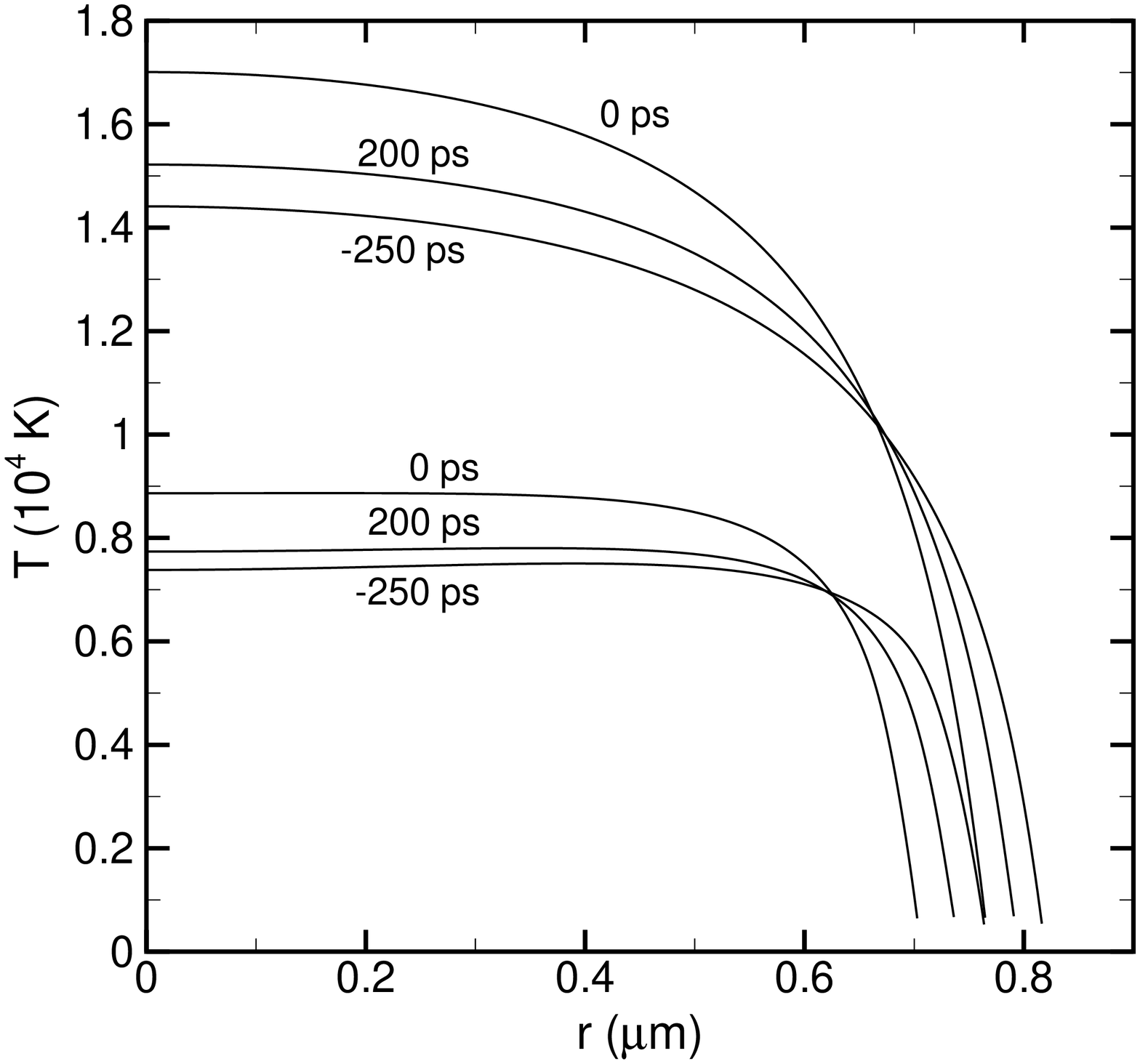}
\end{minipage}
\caption{Snapshots of the temperature distributions for Ar bubble
at $P_a=1.2$~bar, $R_0=4.5~\mu$m. The upper three lines are for
nonreacting case II,  and the lower three lines are for reacting
case III.  $t=0$ ps corresponds to the time of minimum radius
($t_\text{min}=19.015945~ \mu$s).} \label{fig2}
\end{figure}

\begin{figure}[]
\begin{minipage}[t]{0.49\textwidth}
\includegraphics[width= 8cm ]{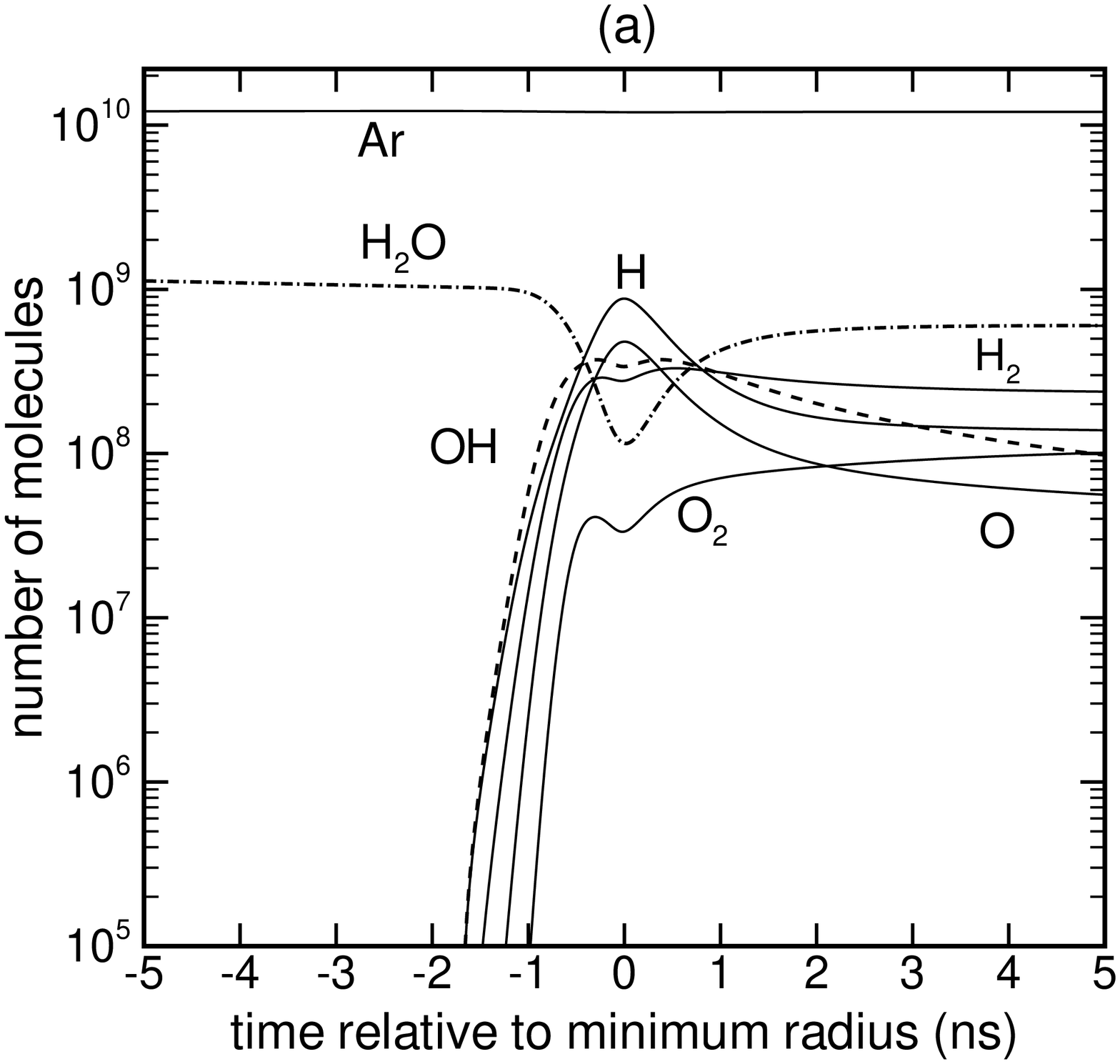}
\end{minipage}
\begin{minipage}[t]{0.49\textwidth}
\includegraphics[width= 8cm ]{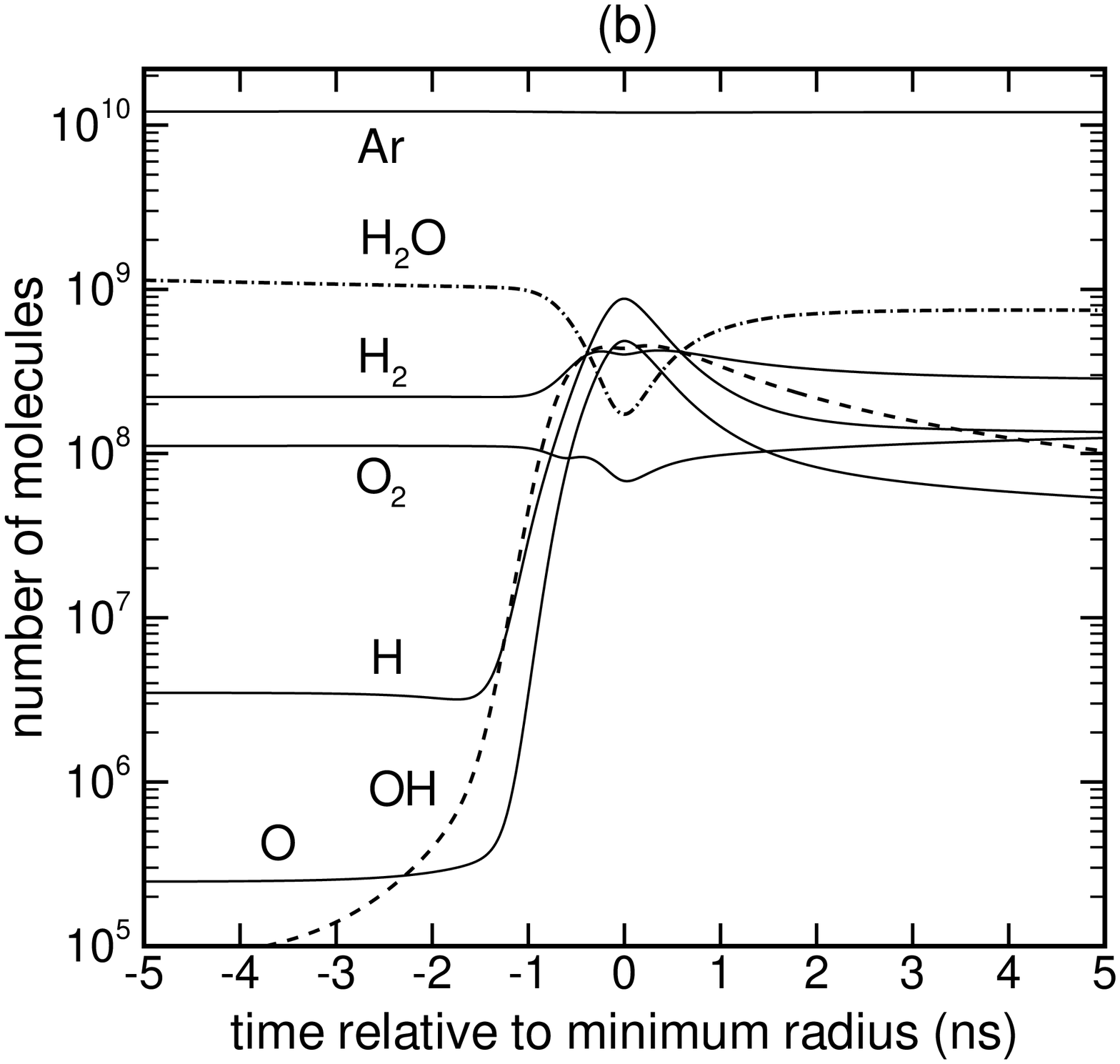}
\end{minipage}

\begin{minipage}[t]{0.49\textwidth}
\includegraphics[width= 8cm ]{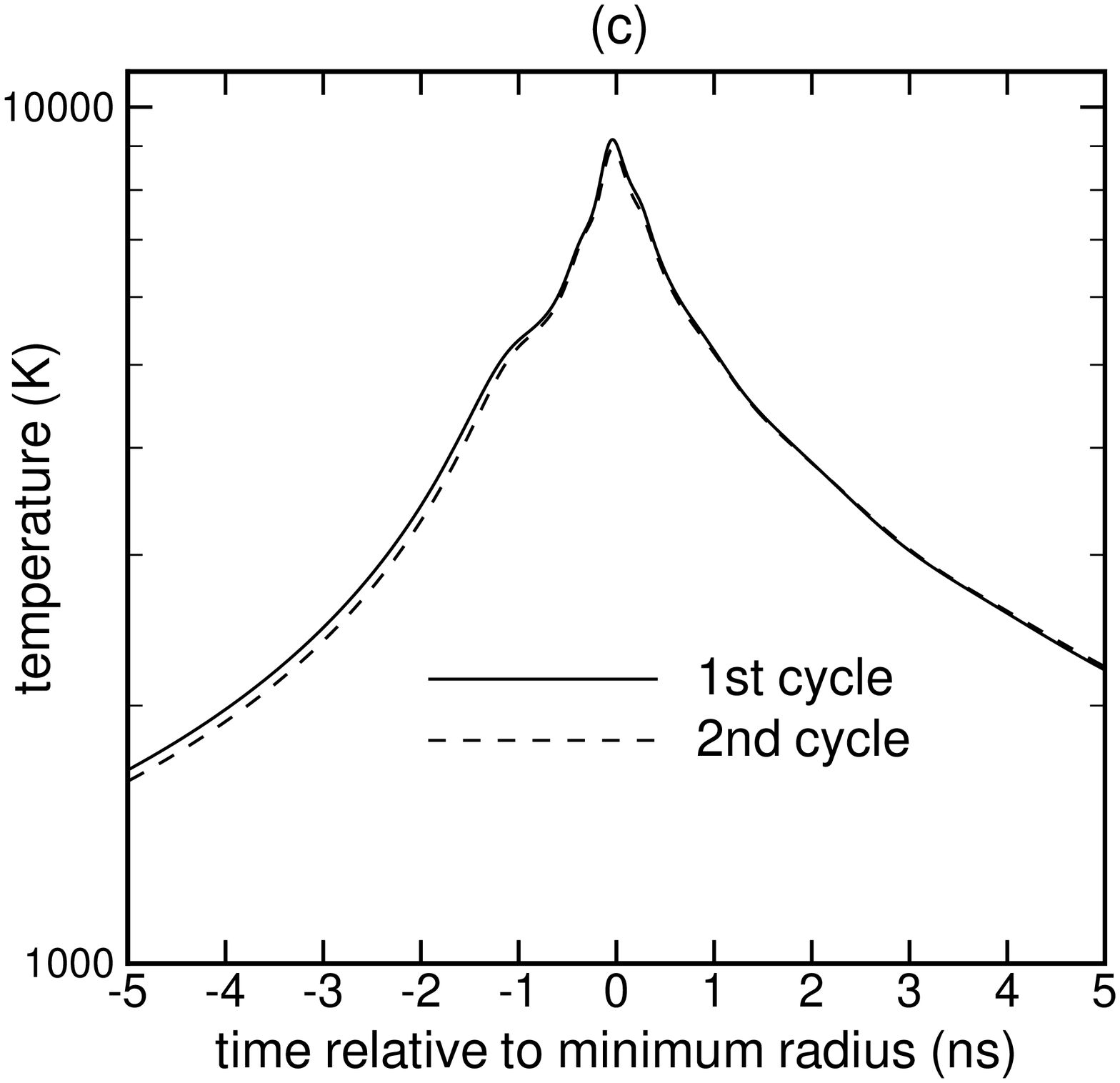}
\end{minipage}\par
\caption{The numbers of molecules of species and the temperature
of the center around the moment of  minimum bubble radius as a
function of time for Ar  bubble at  $P_a=1.2$~bar, $R_0=4.5~\mu$m.
(a) First acoustic cycle, (b) second acoustic cycle, (c)
temperature at the bubble center.} \label{fig3}
\end{figure}

\begin{figure}[]
\begin{minipage}[t]{0.49\textwidth}
\includegraphics[width= 8cm ]{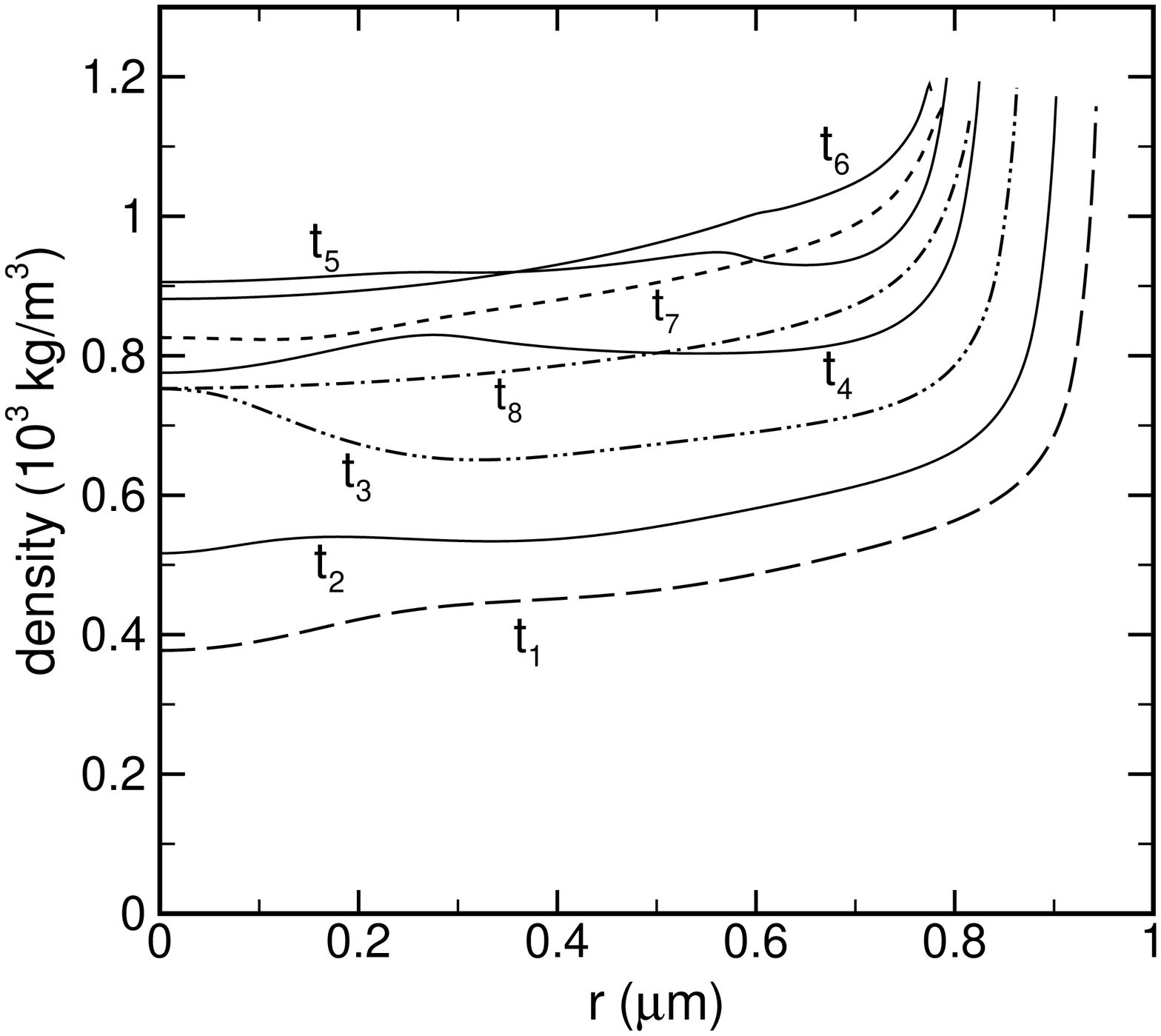}
\end{minipage}
\begin{minipage}[t]{0.49\textwidth}
\includegraphics[width= 8cm ]{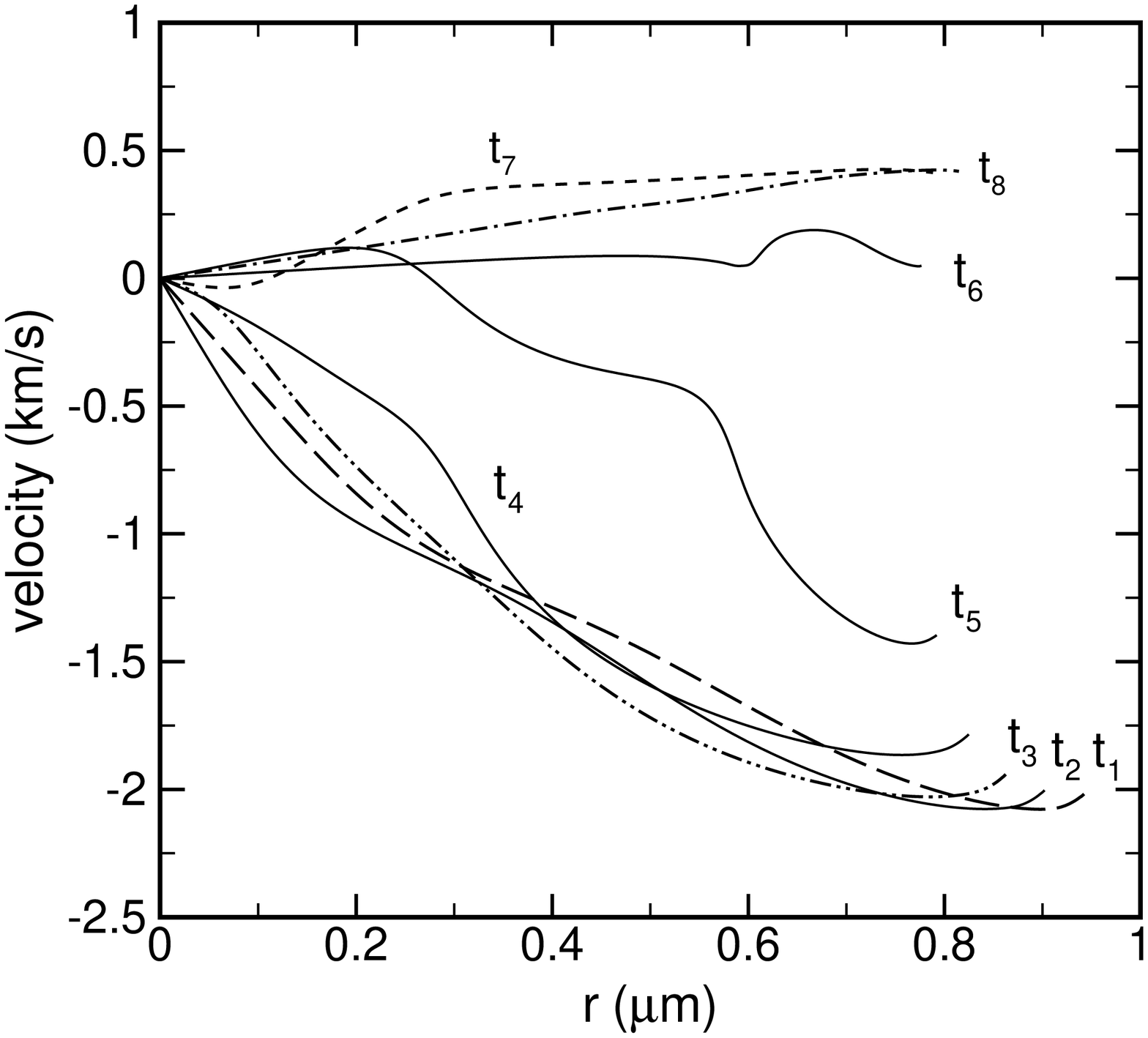}
\end{minipage}

\begin{minipage}[t]{0.49\textwidth}
\includegraphics[width= 8cm ]{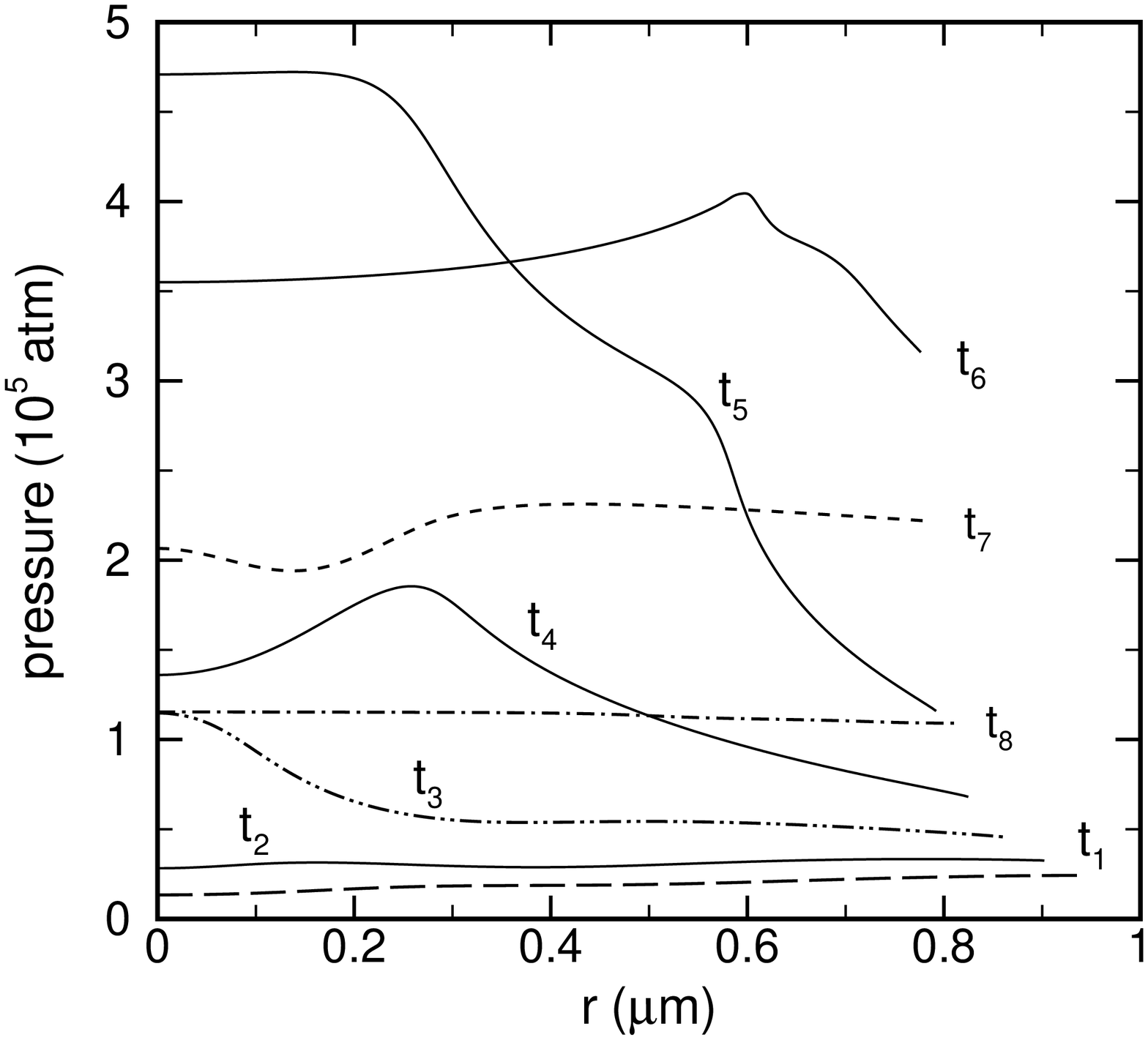}
\end{minipage}
\begin{minipage}[t]{0.49\textwidth}
\includegraphics[width= 8cm ]{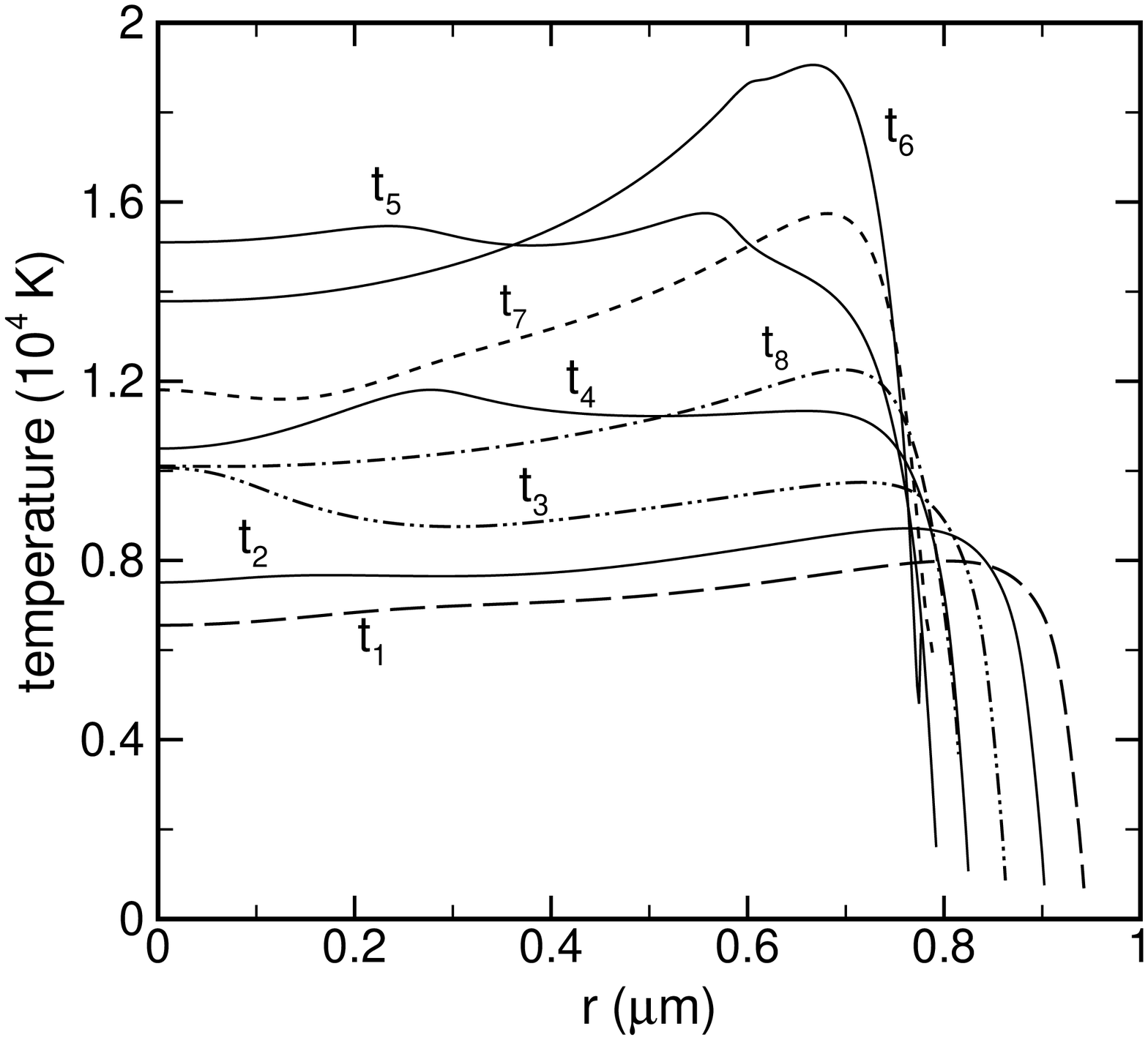}
\end{minipage}\par
\caption{Snapshots of the spatial profiles of density, velocity,
pressure,  and temperature for  $P_a=1.4$~bar, $R_0=6~\mu$m. Time
sequences are $t_1=-100$~ps, $t_2=-80$~ps, $t_3=-60$~ps,
$t_4=-40$~ps, $t_5=-20$~ps, $t_6=0$~ps, $t_7= 40 $~ps,
$t_8=100$~ps, where $t_6=0$ denotes the time of minimum radius
($t_\text{min}=28.970309~ \mu$s).} \label{fig4}
\end{figure}

\begin{figure}[]
\begin{minipage}[t]{0.49\textwidth}
\includegraphics[width= 8cm ]{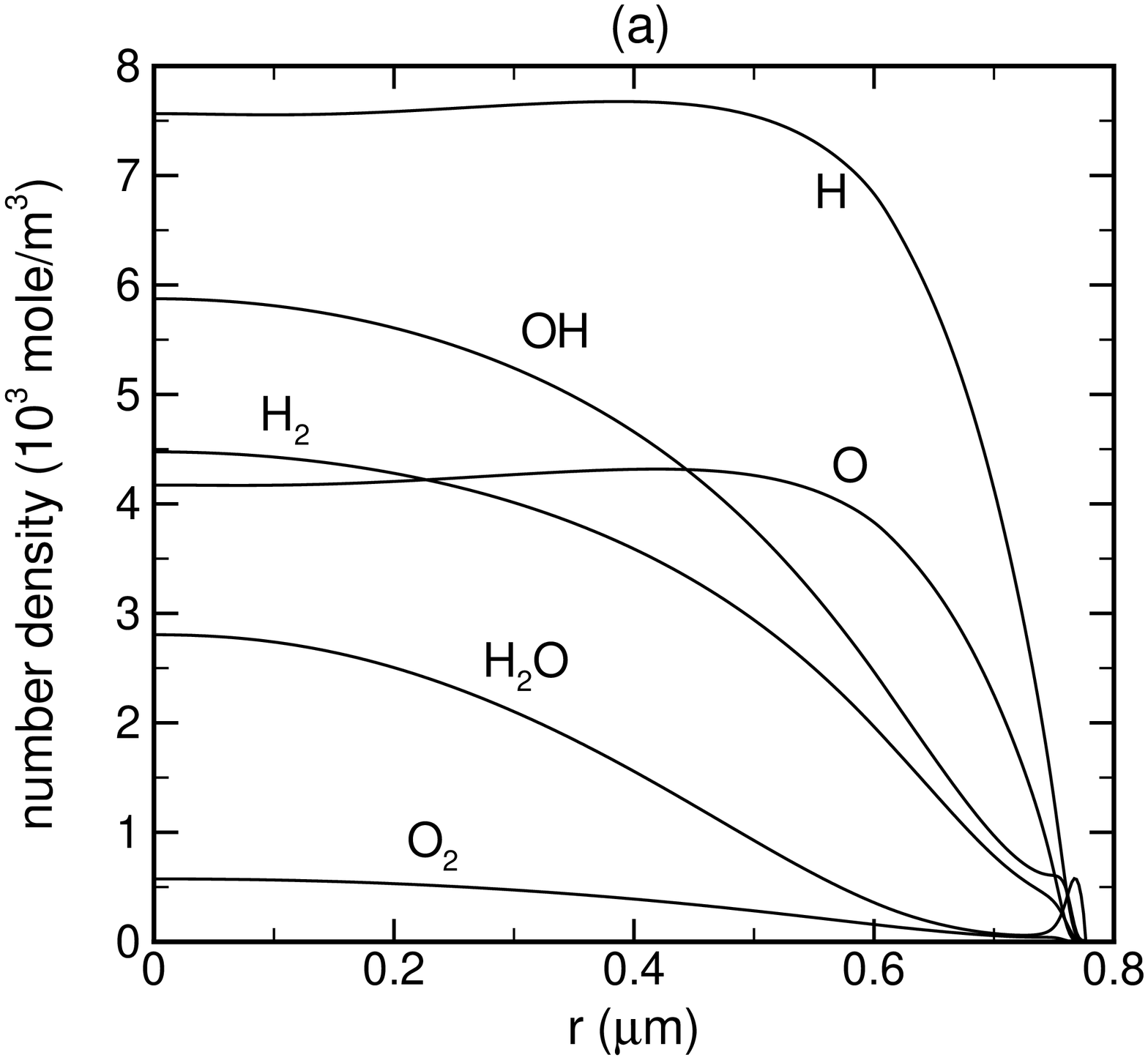}
\end{minipage}
\begin{minipage}[t]{0.49\textwidth}
\includegraphics[width= 8cm ]{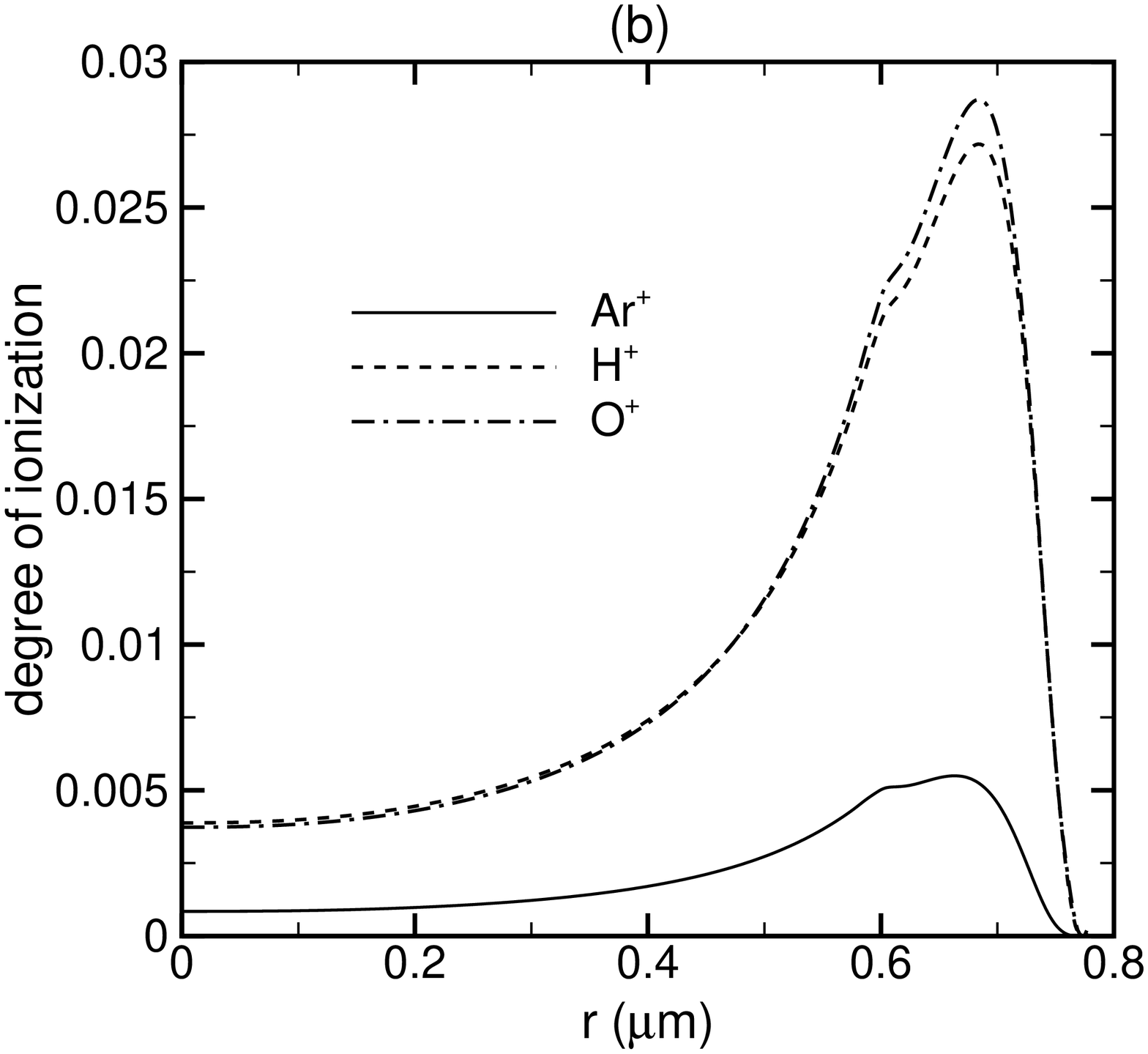}
\end{minipage}
\caption{The spatial profiles of number densities for  molecular
species (a) and degrees of ionization for ionic species (b) at the
time of minimum bubble radius $t=0$ ($t_\text{min}=28.970309~
\mu$s) for $P_a=1.4$~bar, $R_0=6~\mu$m. The degree of ionization
is computed  using the Saha equation.} \label{fig5}
\end{figure}

\begin{figure}[]
\begin{minipage}[t]{0.49\textwidth}
\includegraphics[width= 8cm ]{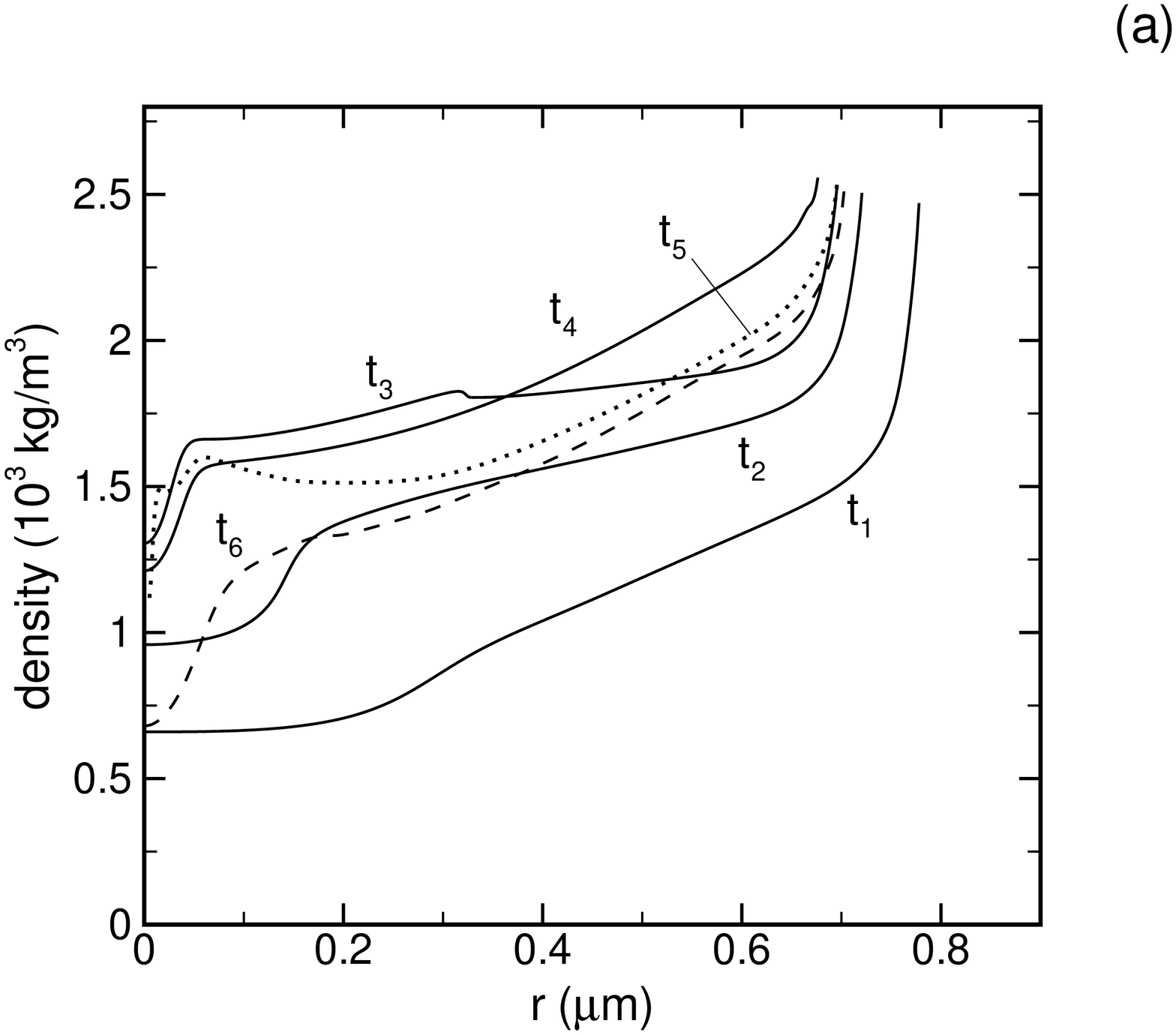}
\end{minipage}
\begin{minipage}[t]{0.49\textwidth}
\includegraphics[width= 8cm ]{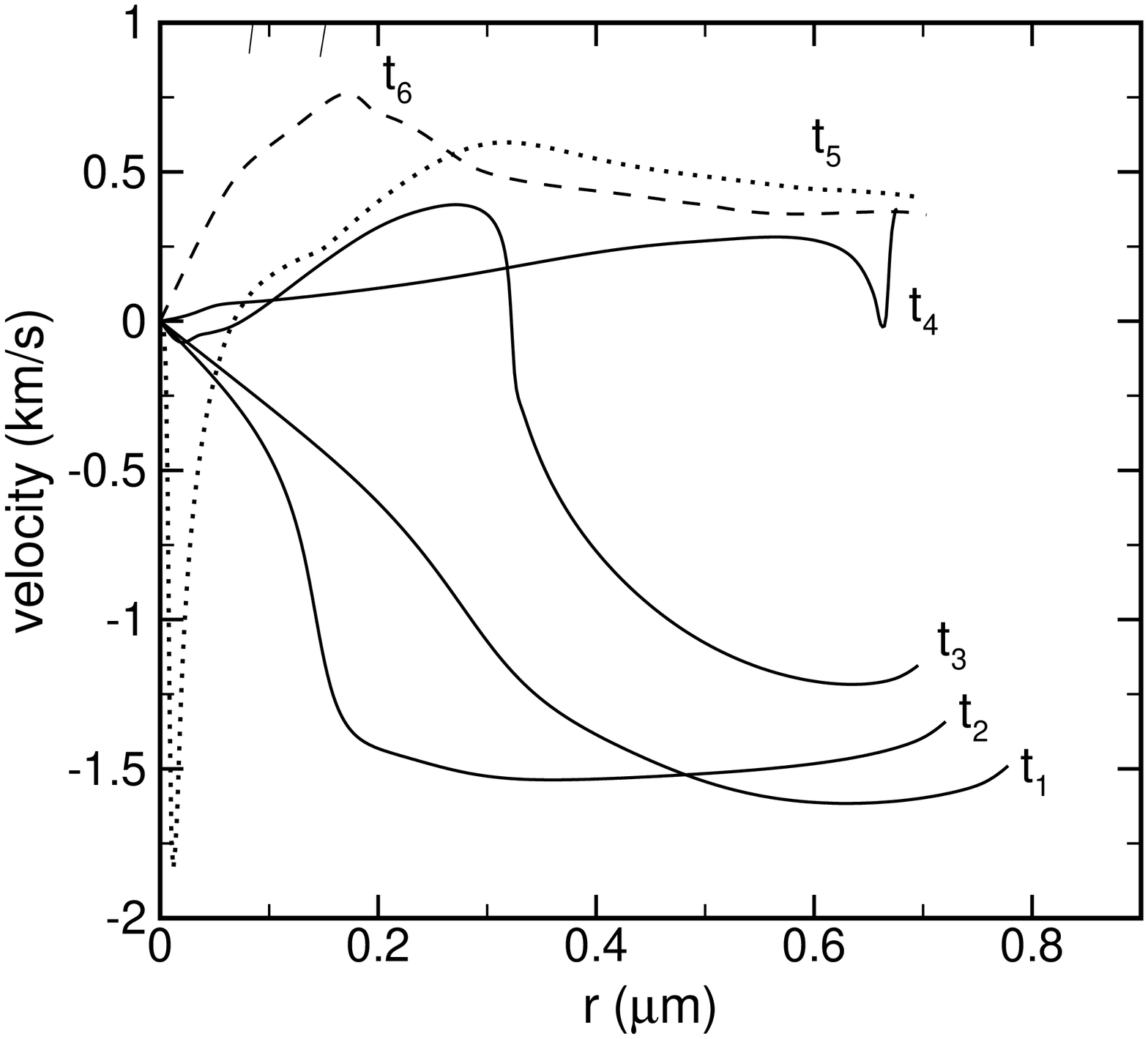}
\end{minipage}

\begin{minipage}[t]{0.49\textwidth}
\includegraphics[width= 8cm ]{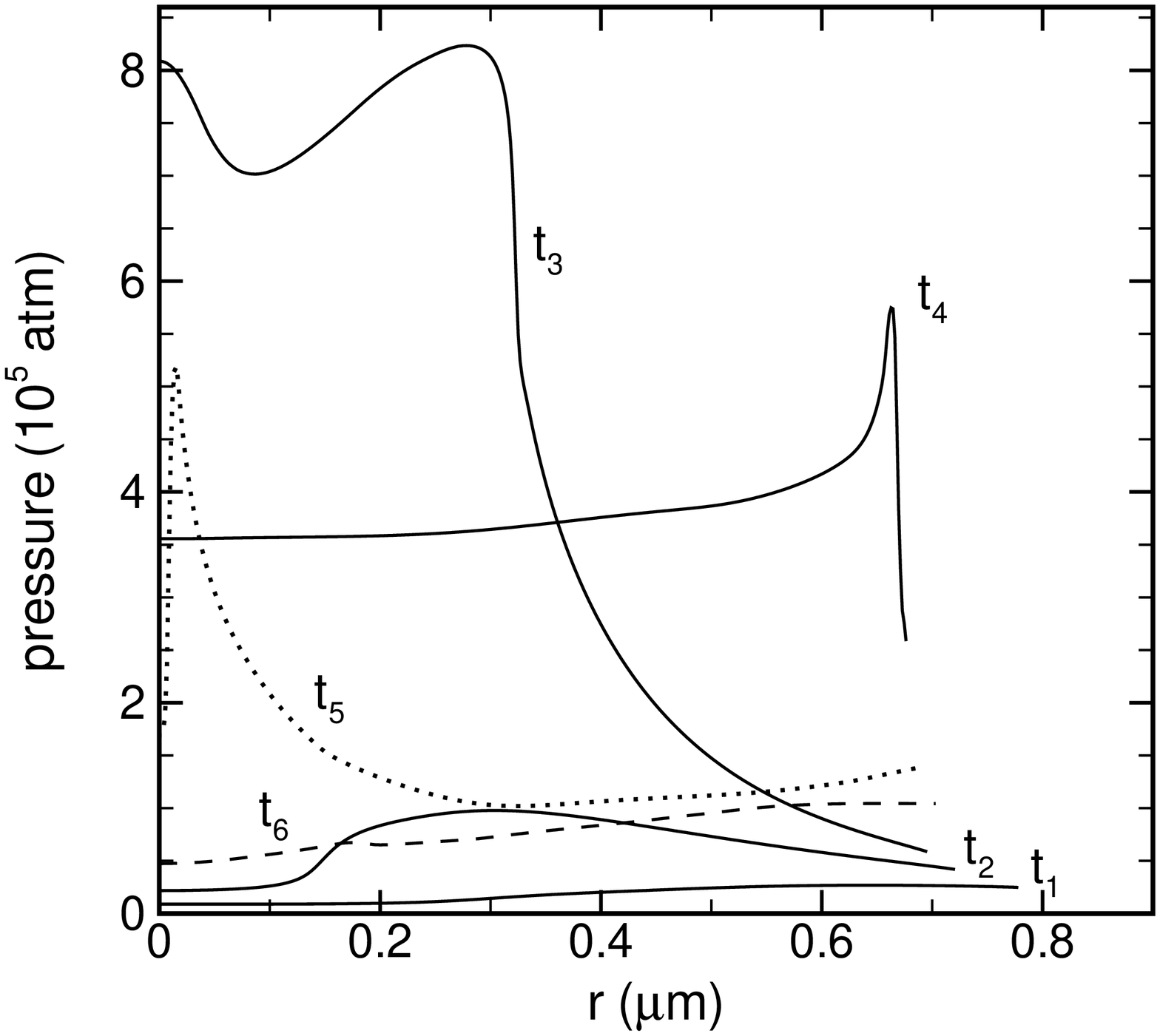}
\end{minipage}
\begin{minipage}[t]{0.49\textwidth}
\includegraphics[width= 8cm ]{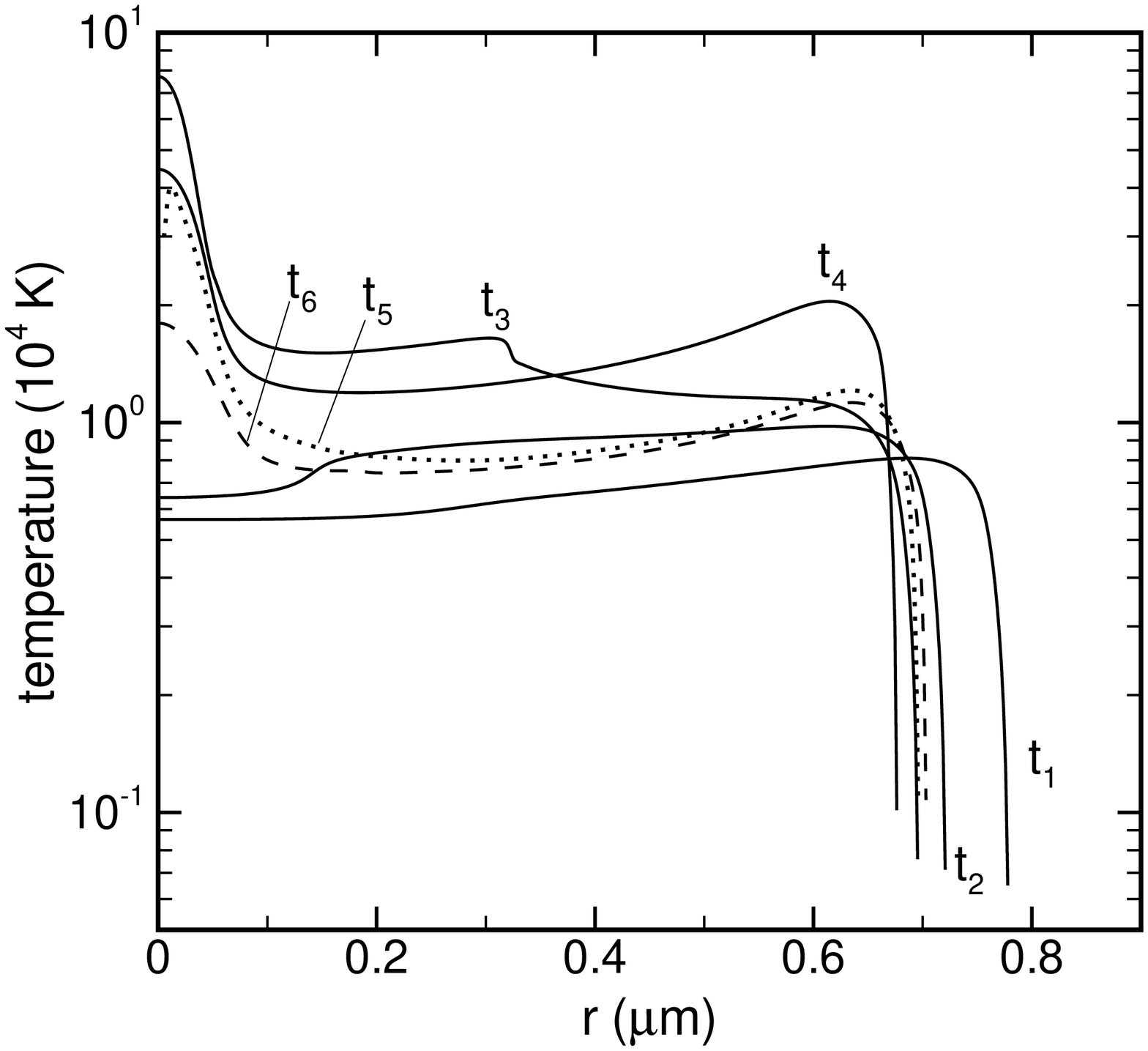}
\end{minipage}
\end{figure}

\begin{figure}[]
\begin{minipage}[t]{0.49\textwidth}
\includegraphics[width= 8cm ]{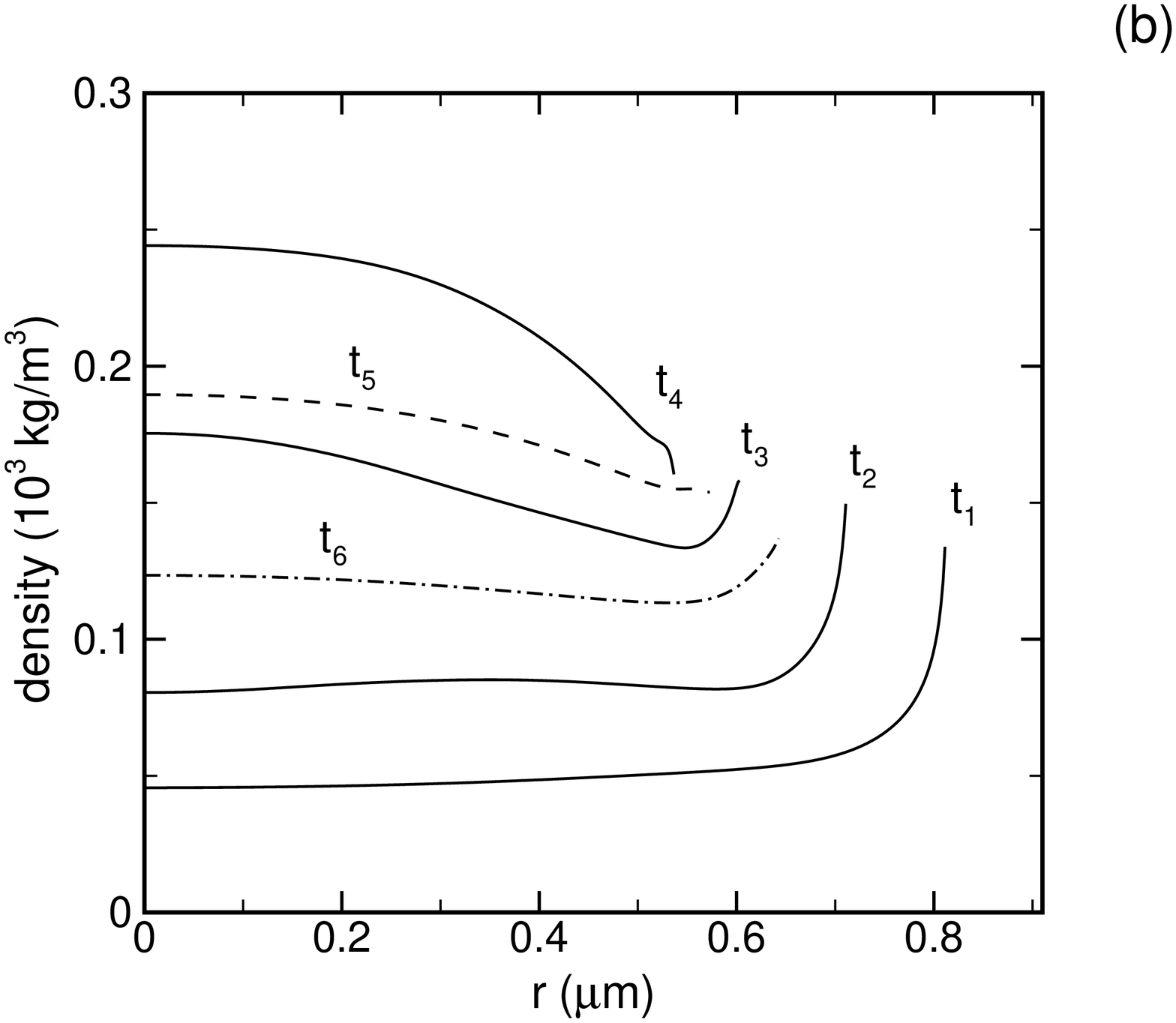}
\end{minipage}
\begin{minipage}[t]{0.49\textwidth}
\includegraphics[width= 8cm ]{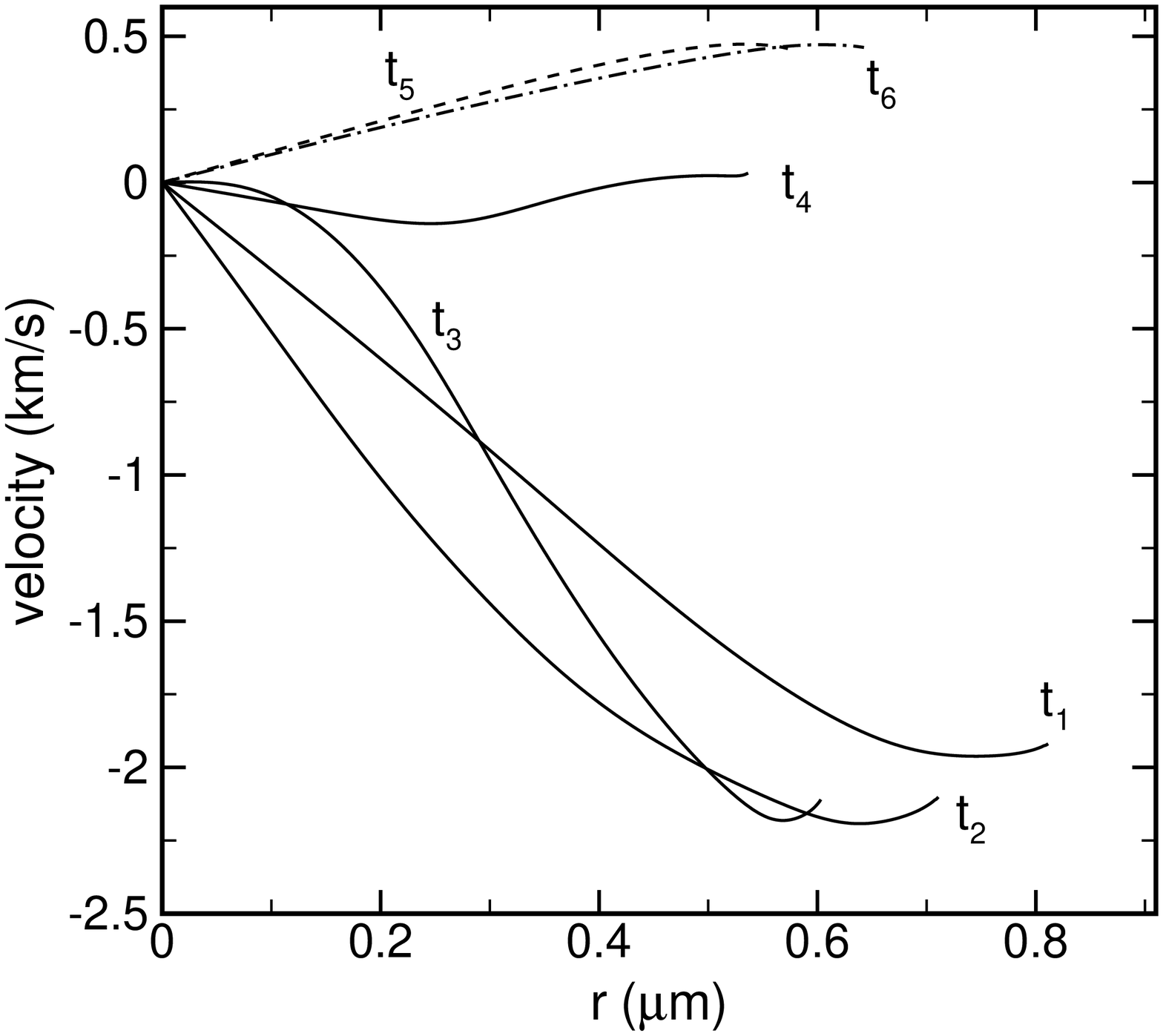}
\end{minipage}

\begin{minipage}[t]{0.49\textwidth}
\includegraphics[width= 8cm ]{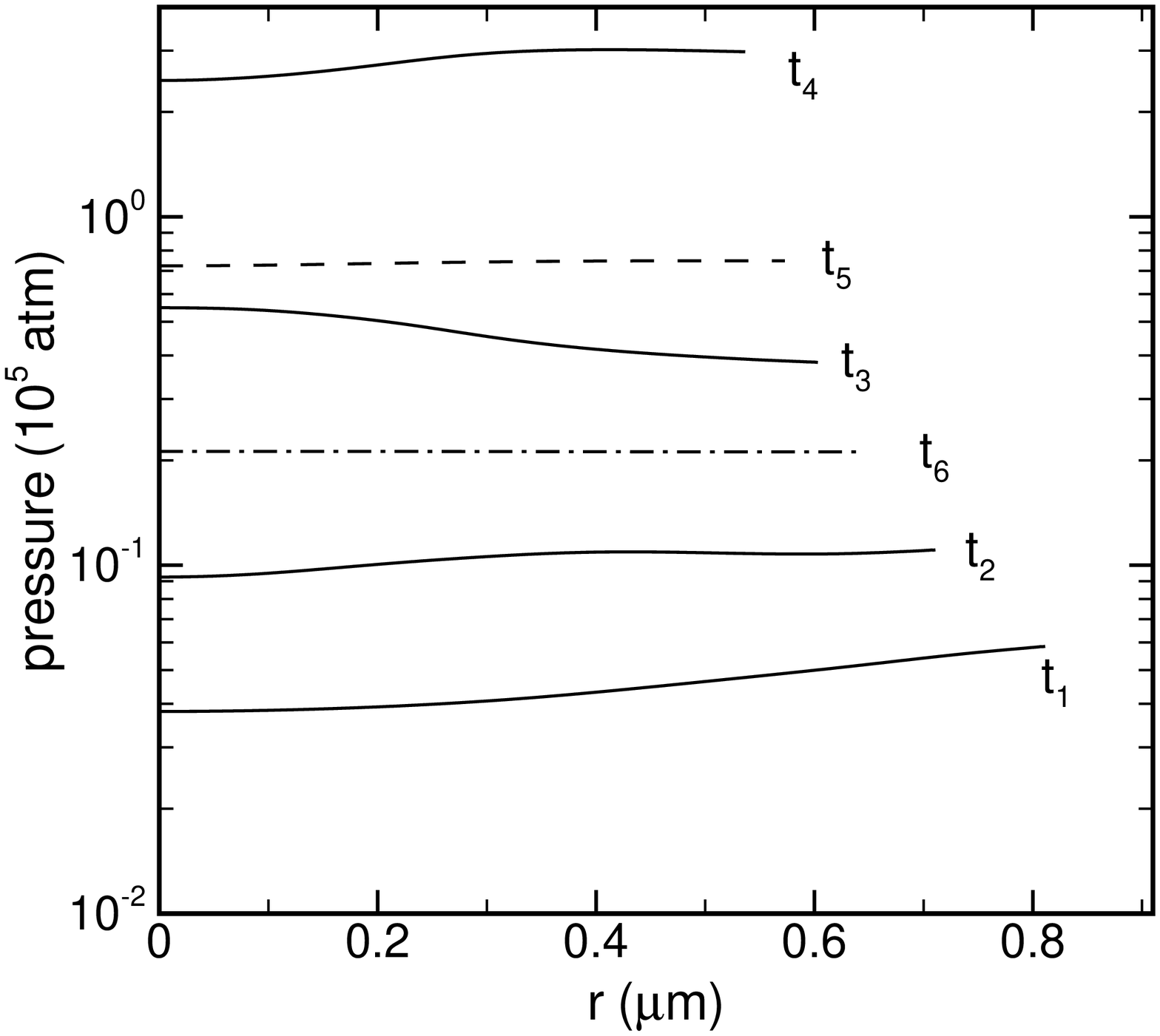}
\end{minipage}
\begin{minipage}[t]{0.49\textwidth}
\includegraphics[width= 8cm ]{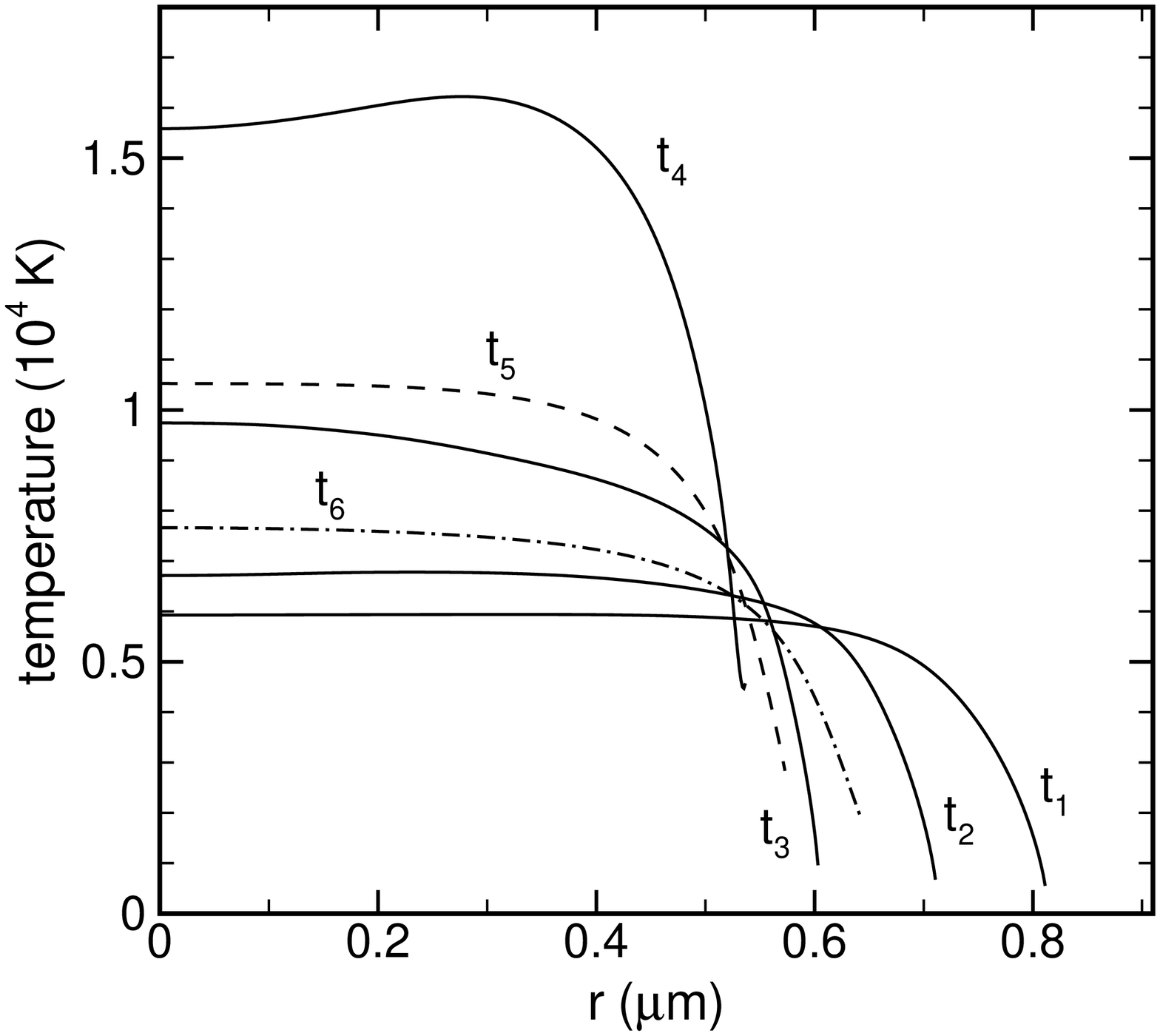}
\end{minipage}
\caption{ Snapshots of the spatial profiles of density, velocity,
pressure and temperature for  $P_a=1.35$~atm, $R_0=4.5~\mu$m. (a)
Xe bubble. Time sequences are $t_1=-80$~ps, $t_2=-40$~ps,
$t_3=-20$~ps, $t_4=0$~ps, $t_5=40~$ps, $t_6=60~$ps, where $t_4=0$
denotes the time of minimum radius
($t_\text{min}=22.006191~\mu$s). (b) He bubble. $t_1=-150~$ps,
$t_2=-100~$ps, $t_3=-50~$ps, $t_4=0~$ps, $t_5=100~$ps,
$t_6=250~$ps, where $t_\text{min}=22.057459~\mu$s. } \label{fig6}
\end{figure}

\begin{figure}
\begin{minipage}[t]{0.49\textwidth}
\includegraphics[width= 8cm ]{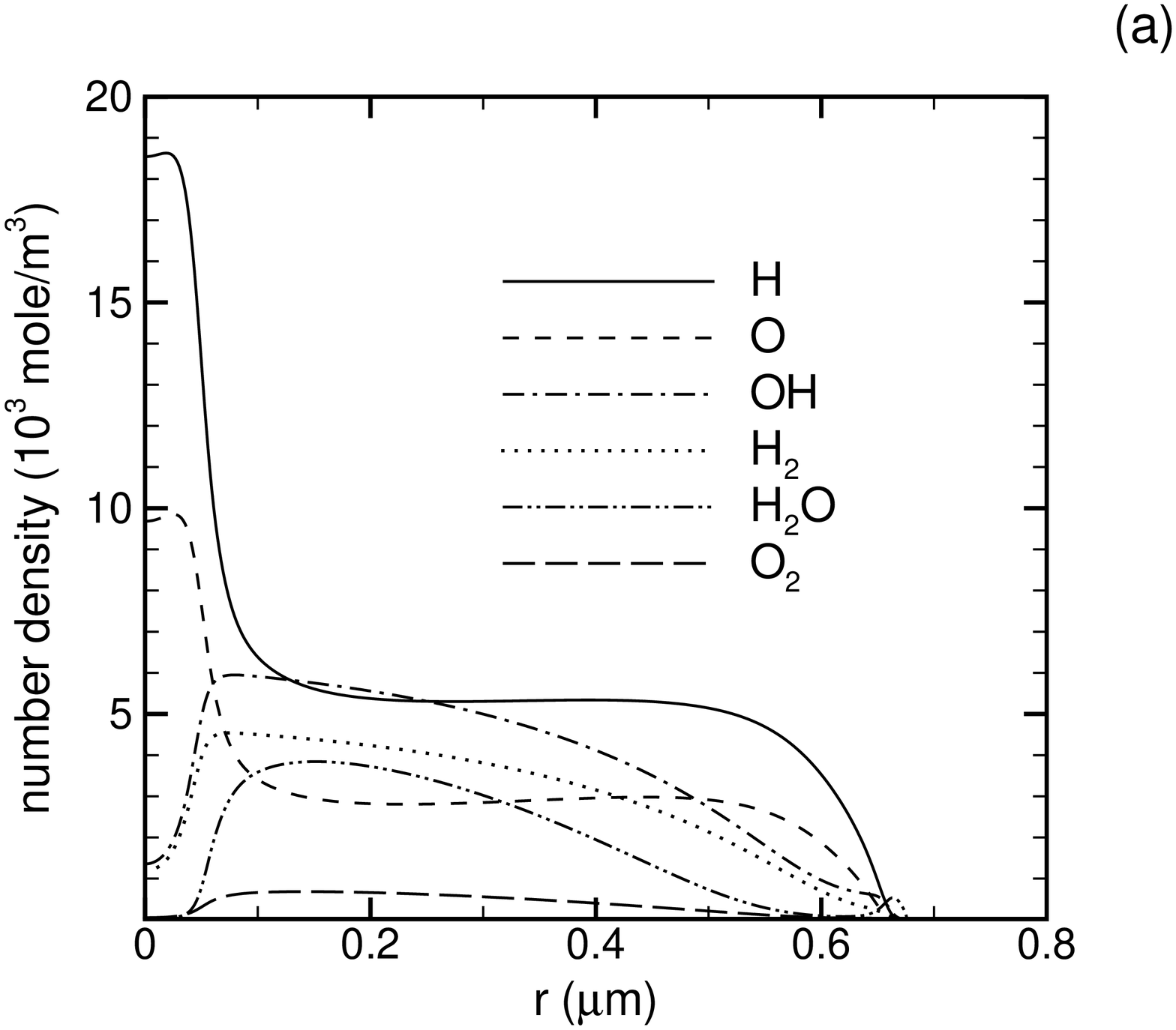}
\end{minipage}
\begin{minipage}[t]{0.49\textwidth}
\includegraphics[width= 8cm ]{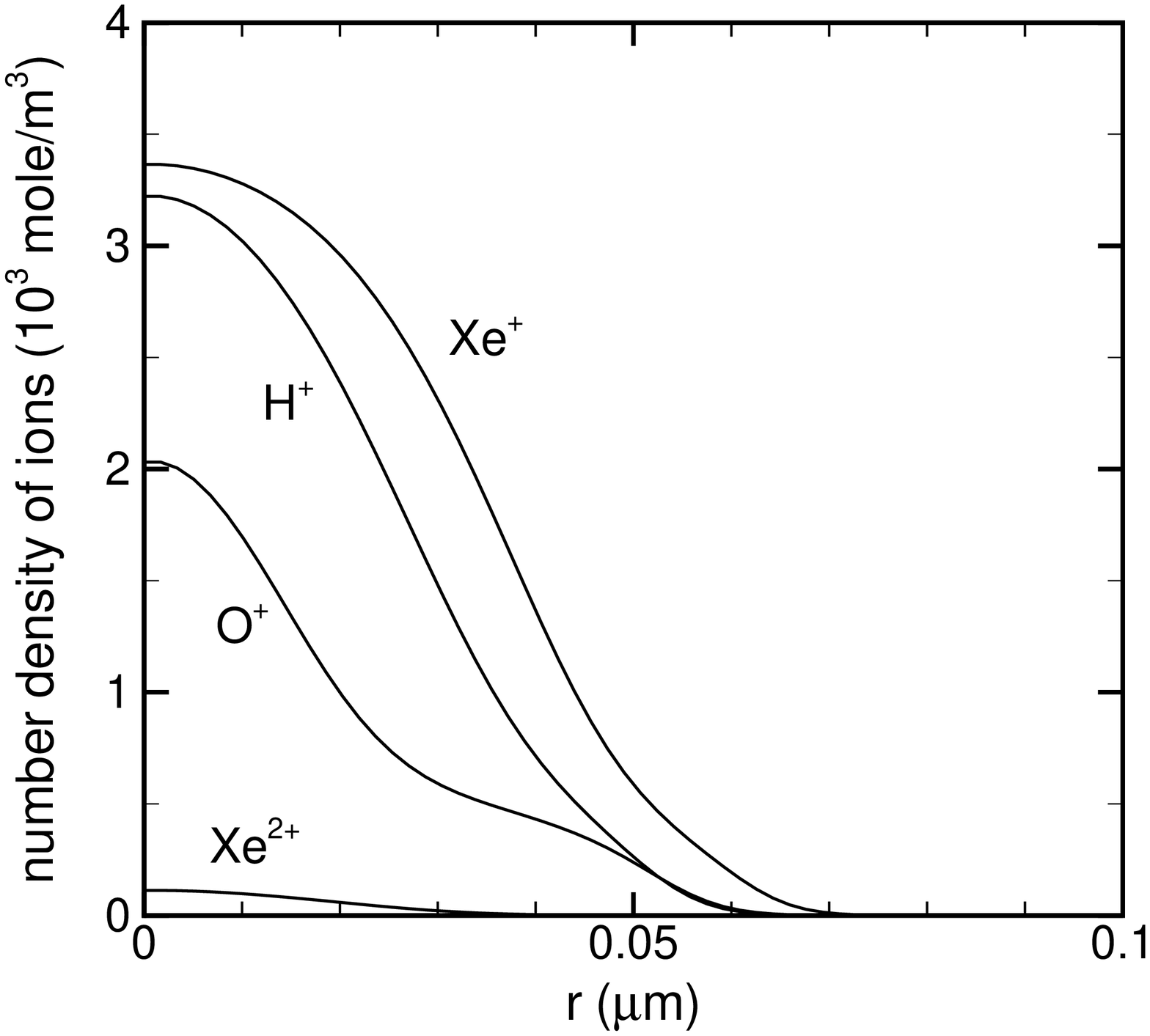}
\end{minipage}

\begin{minipage}[t]{0.49\textwidth}
\includegraphics[width= 8cm ]{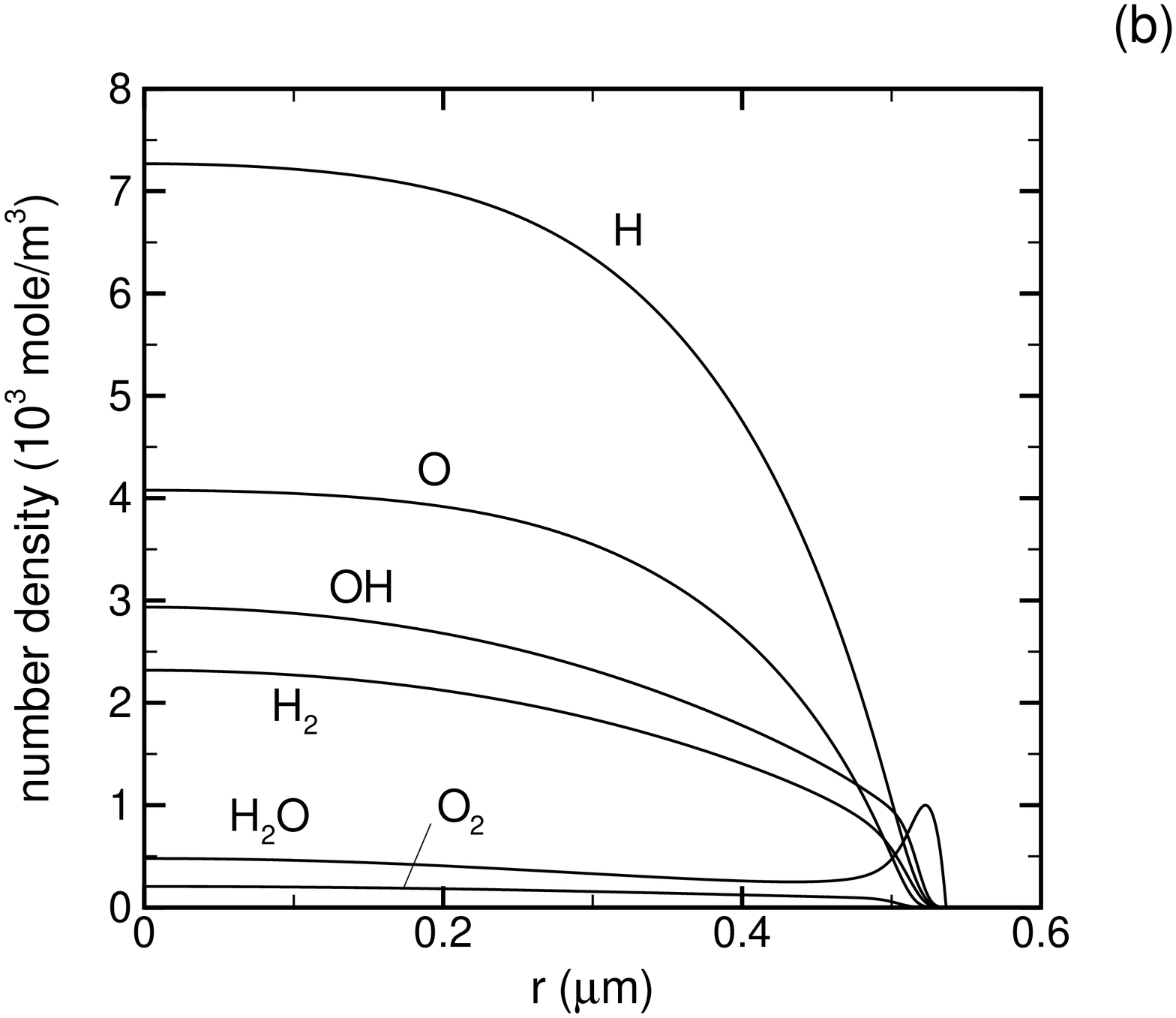}
\end{minipage}
\begin{minipage}[t]{0.49\textwidth}
\includegraphics[width= 8cm ]{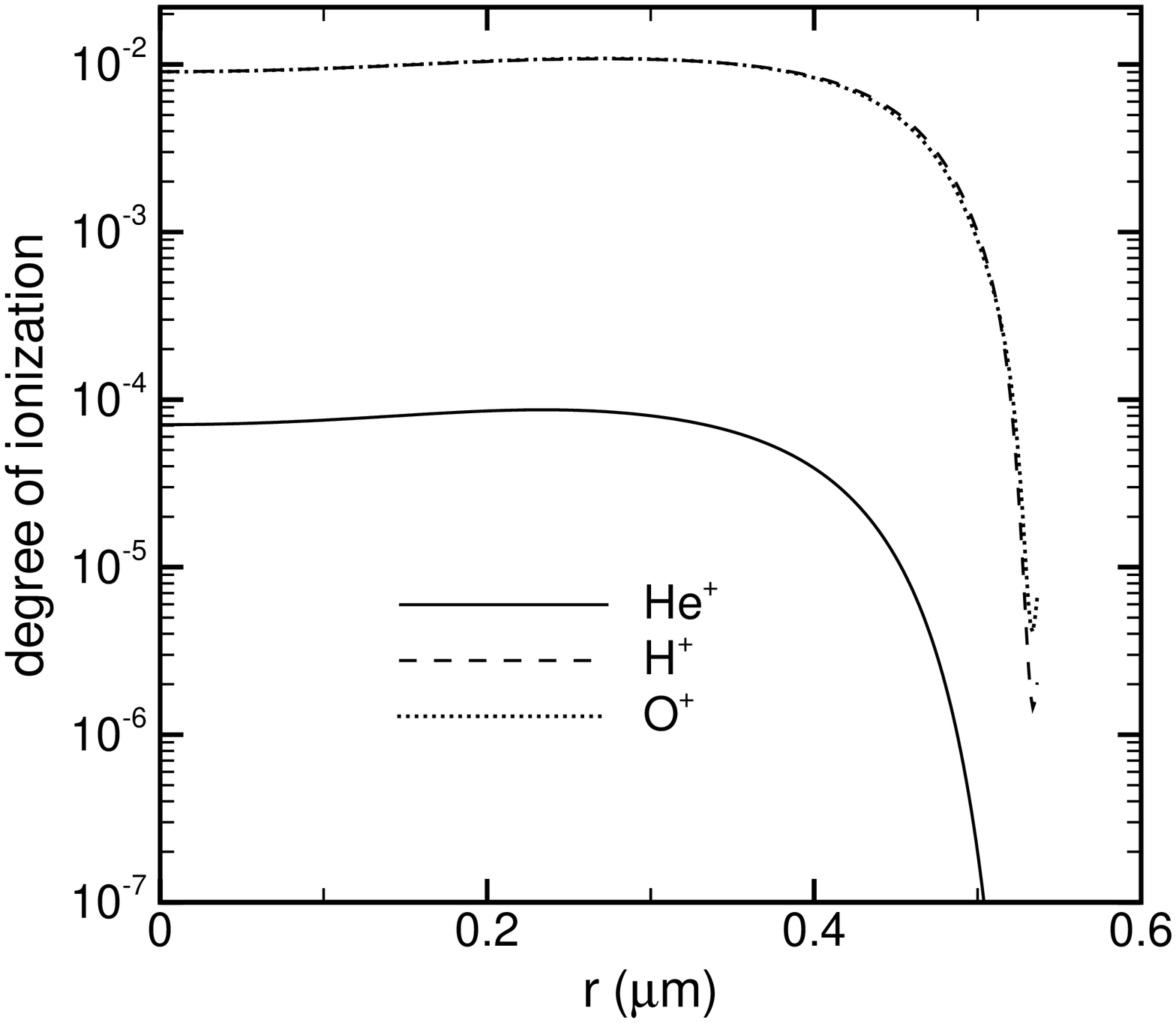}
\end{minipage}
\caption{The spatial profiles of the number densities of molecular
species (left) and the compositions of ions (right) for
$P_a=1.35$~atm, $R_0=4.5~\mu$m. (a) Xe bubble with ions computed
using the nonequilibrium ionization. (b) He bubble with the degree
of ionization computed using the Saha equation. All species are
shown for the time of minimum radius. }
\end{figure}

\begin{figure}
\includegraphics[width= 8cm ]{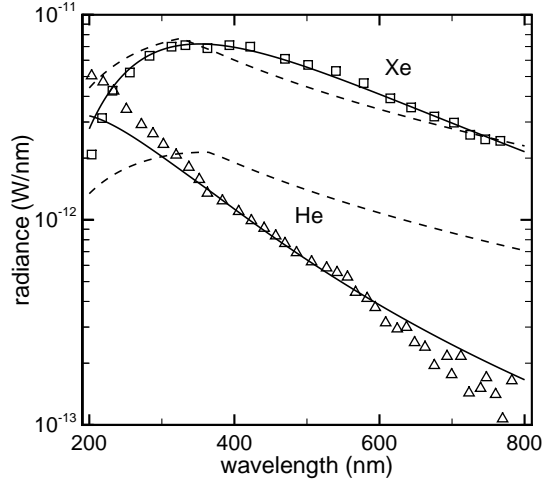}
\caption{Spectral radiance of the SL light from bubbles of Xe and
He in water. The same parameters as in experiment \cite{Vaz01} are
Xe (the ambient radius $R_0=5.5~\mu$m, dissolved partial pressure
3 Torr) and He ($R_0=4.5~\mu$m, 150 Torr), water temperature
23~$^\circ$C, driving frequency 42 kHz.  The squares and triangles
are experimental spectra of Xe and He bubbles, respectively. The
solid lines are calculated spectra of the \emph{ad hoc}
finite-size blackbody model with fitting parameters
$P_a=1.28~$atm, $E_c=1.80\times 10^{4}$ (Xe), and $P_a=1.45~$atm,
$E_c=2.50 \times 10^{2}$ (He), and the dashed lines are those of
the optically thin model with fitting parameters $P_a=1.55~$atm
(Xe) and $2.0~$atm (He).} \label{fig8}
\end{figure}

\begin{figure}
\begin{minipage}[t]{0.49\textwidth}
\includegraphics[width= 8cm ]{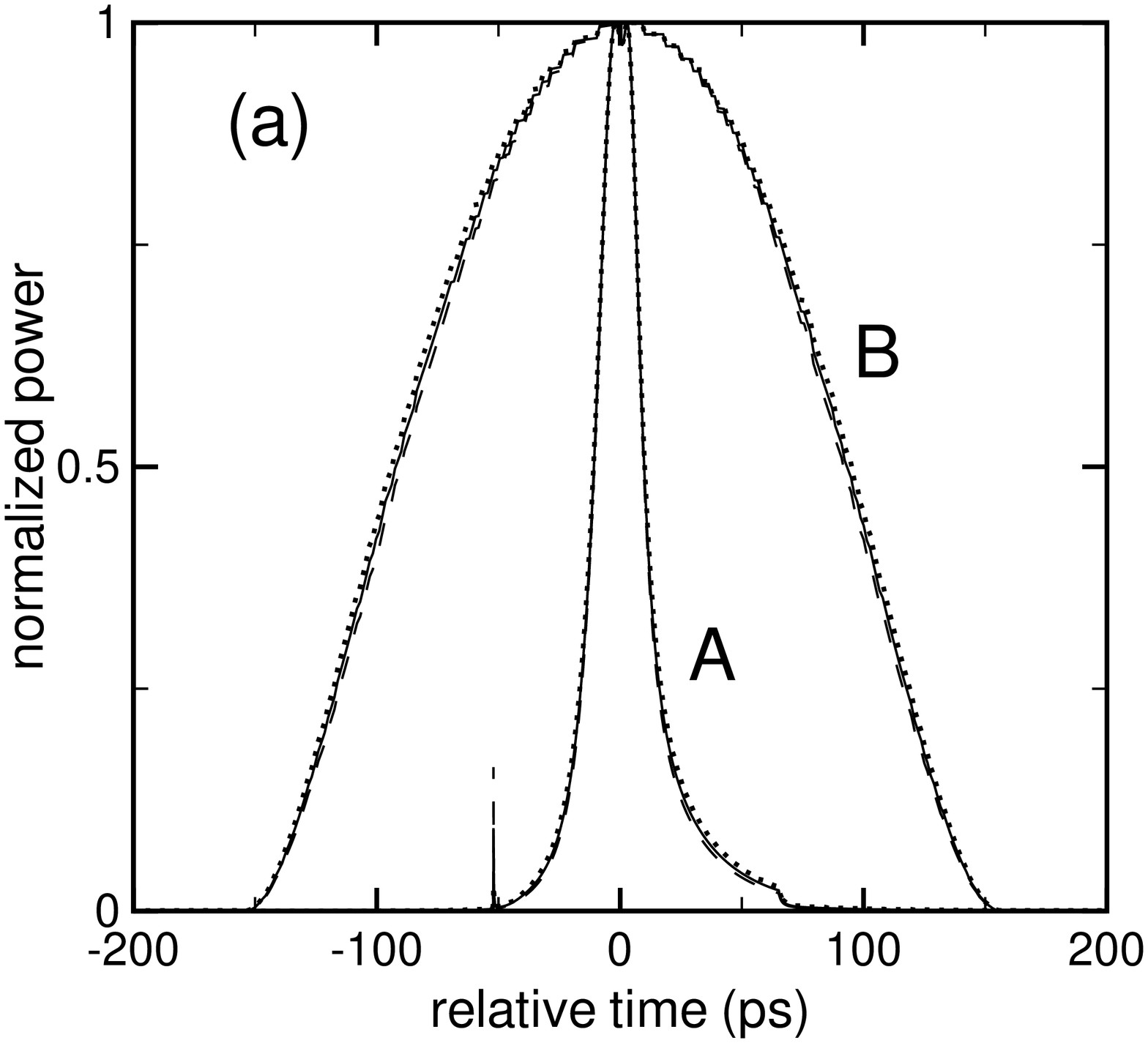}
\end{minipage}
\begin{minipage}[t]{0.49\textwidth}
\includegraphics[width= 8cm ]{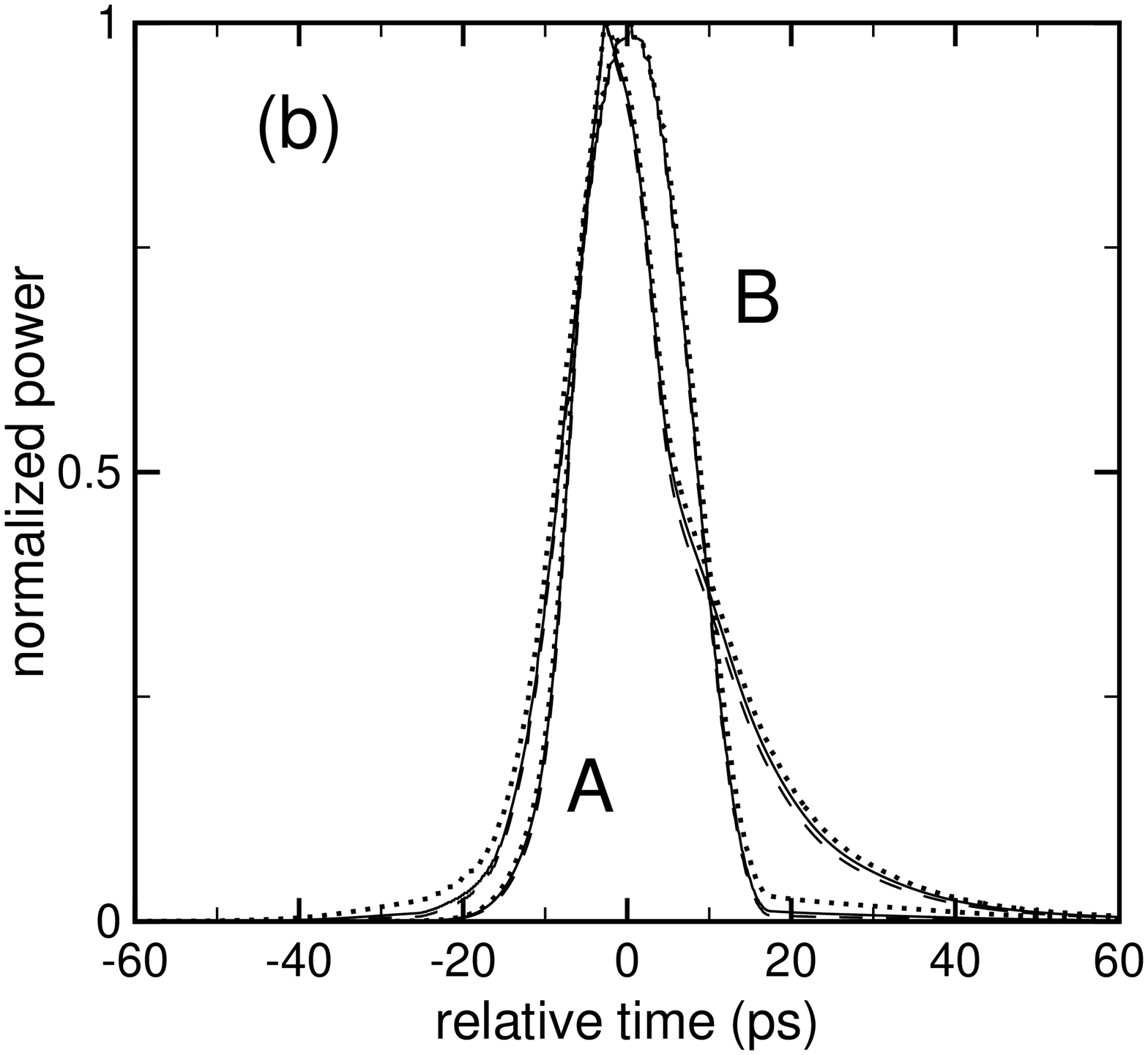}
\end{minipage}
\caption{Normalized radiation power vs time for the optically thin
model (``A") and  the \emph{ad hoc} finite-size blackbody model
(``B") in Xe bubble (a) and He bubble (b). The solid line denotes
the total measurable power, the dashed line in the UV range (300
nm $< \lambda < 400~$nm), and the dotted line in the red range
(590 nm $< \lambda < 650~$nm). Time is relative to the moment of
minimum bubble radius. The parameters are the same as in Fig.~8.}
\label{fig9}
\end{figure}

\begin{figure}
\includegraphics[width= 8cm ]{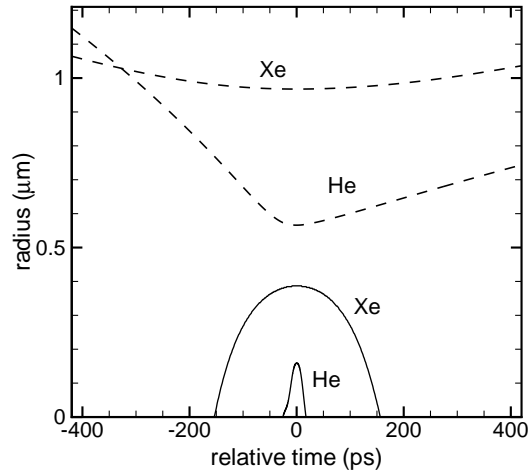}
\caption{The blackbody core $R_c$ (solid line) and the bubble
radius $R$ (dashed line) vs time for Xe and He bubbles. The
parameters are the same as those for the \emph{ad hoc} finite-size
blackbody model in Fig.~8.} \label{fig10}
\end{figure}

\begin{figure}
\includegraphics[width= 16cm ]{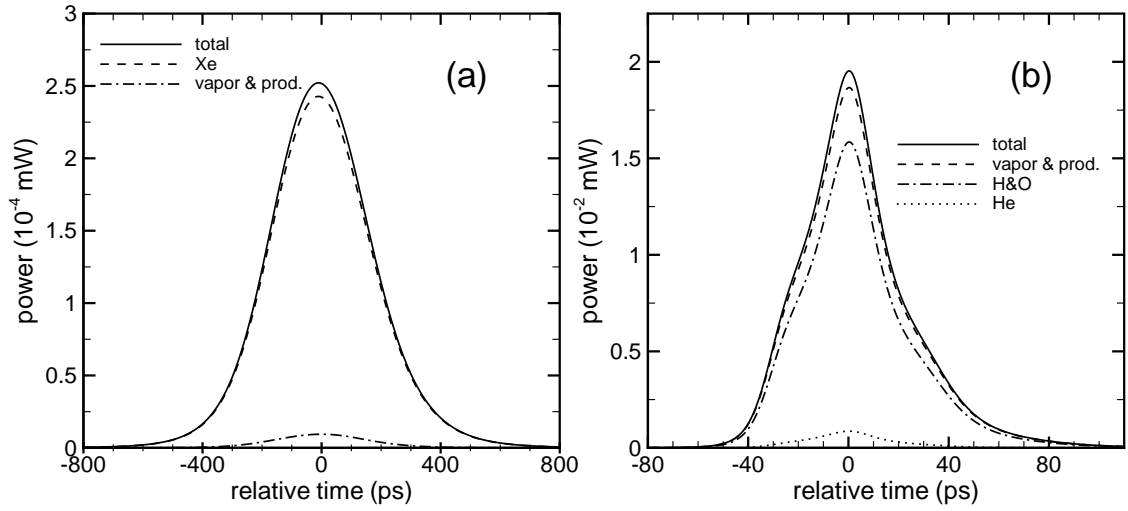}
\caption{The optical powers  vs time for Xe (a) and He (b) bubbles
computed  from the optically thin model. The uppermost curve is
the total power, and others are powers contributed from the
species as marked. The driving pressure amplitudes are $1.28$~atm
for  Xe bubble, and $1.45$~atm for  He bubble. Other parameters
are the same as those in Fig.~8. Time is relative to the moment of
minimum bubble radius.} \label{fig11}
\end{figure}

\end{document}